\documentclass[twocolumn,epjc3]{svjour3}
\usepackage[utf8]{inputenc}
\usepackage[T1]{fontenc}
\usepackage[english]{babel}

\usepackage[numbers,sort&compress]{natbib}

\usepackage{graphicx}
\usepackage{scalefnt}

\usepackage{amsmath}
\usepackage{amssymb}
\usepackage{amsfonts}
\usepackage{mathtools}
\usepackage{braket}
\usepackage[force]{feynmp-auto}

\usepackage{url}
\usepackage{xurl}

\usepackage[dvipsnames]{xcolor}

\usepackage{txfonts}
\usepackage{microtype}

\usepackage{xspace} 

\usepackage{hyphenat}
\usepackage[anythingbreaks]{breakurl}

\usepackage[breaklinks=true]{hyperref}

\hypersetup{
    colorlinks = true,
    citecolor  = blue,
    linkcolor  = blue,
    urlcolor  = blue
}

\newcommand{\exchange}{[\eta \leftrightarrow \Omega]}

\newcommand{\la}{\langle}
\newcommand{\ra}{\rangle}

\newcommand{\opno}[3]{\ensuremath{ {#1}^{#2}_{#3}}}

\newcommand{\permtwo}[2]{P(#1/#2)}

\newcommand{\opA}{\ensuremath{A}}
\newcommand{\opB}{\ensuremath{B}}
\newcommand{\opC}{\ensuremath{C}}

\newcommand{\qpcre}[1]{\beta^\dagger_{#1}}
\newcommand{\qpan}[1]{\beta_{#1}}
\newcommand{\ptorder}[1]{\mathcal{O}(\lambda^{#1})}

\newcommand{\dds}[1]{\frac{\mathrm{d} #1}{\mathrm{d} s}}
\newcommand{\magnus}{M}

\journalname{Eur. Phys. J. A}
\renewcommand{\email}[1]{e-mail: \href{mailto:#1}{#1}}

\begin{document}

\allowdisplaybreaks

\title{ADG: Automated generation and evaluation of many-body diagrams}
\subtitle{III. Bogoliubov in-medium similarity renormalization group formalism}

\author{
A.~Tichai\thanksref{ad:tud,ad:emmi,ad:mpik,em:at} \and 
P. Arthuis\thanksref{ad:tud,ad:emmi,em:pa} \and 
H.~Hergert\thanksref{ad:msu,em:hh} \and
T.~Duguet\thanksref{ad:irfu,ad:leuven,em:td}
}
\date{Received: \today{} / Accepted: date}

\thankstext{em:at}{\email{alexander.tichai@physik.tu-darmstadt.de}}
\thankstext{em:pa}{\email{parthuis@theorie.ikp.physik.tu-darmstadt.de}}
\thankstext{em:hh}{\email{hergert@frib.msu.edu}}
\thankstext{em:td}{\email{thomas.duguet@cea.fr}}

\institute{%
\label{ad:tud}%
Technische Universit\"at Darmstadt, Department of Physics, 64289 Darmstadt, Germany
\and 
\label{ad:emmi}%
ExtreMe Matter Institute EMMI, GSI Helmholtzzentrum f\"ur Schwerionenforschung GmbH, 64291 Darmstadt, Germany
\and 
\label{ad:mpik}%
Max-Planck-Institut f\"ur Kernphysik, Saupfercheckweg 1, 69117 Heidelberg, Germany
\and
\label{ad:msu}%
NSCL/FRIB Laboratory and Department of Physics and Astronomy, Michigan State University, East Lansing, MI 48824-1321, USA
\and
\label{ad:irfu}%
IRFU, CEA, Universit\'e Paris-Saclay, 91191 Gif-sur-Yvette, France
\and
\label{ad:leuven}%
KU Leuven, Instituut voor Kern- en Stralingsfysica, 3001 Leuven, Belgium
}

\maketitle

\begin{abstract}
The goal of the present paper is twofold. 
First, a novel expansion many-body method applicable to superfluid open-shell nuclei, the so-called \emph{Bogoliubov in-medium similarity renormalization group} (BIMSRG) theory, is formulated.
This generalization of standard single-reference IMSRG theory for closed-shell systems parallels the recent extensions of coupled cluster, self-consistent Green's function or many-body perturbation theory. Within the realm of IMSRG theories, BIMSRG provides an interesting alternative to the already existing multi-reference IMSRG (MR-IMSRG) method applicable to open-shell nuclei.

The algebraic equations for low-order approximations, i.e., BIMSRG(1) and BIMSRG(2), can be derived manually without much difficulty.  However, such a methodology becomes already impractical and error prone for the derivation of the BIMSRG(3) equations, which are eventually needed to reach high accuracy. Based on a diagrammatic formulation of BIMSRG theory, the second objective of the present paper is thus to describe the third version (v3.0.0) of the \textbf{\texttt{ADG}} code that automatically (1) generates all valid BIMSRG(n) diagrams and (2) evaluates their algebraic expressions in a matter of seconds.  This is achieved in such a way that equations can easily be retrieved for both the flow equation and the Magnus expansion formulations of BIMSRG.

Expanding on this work, the first future objective is to numerically implement BIMSRG(2) (eventually BIMSRG(3)) equations and perform \textit{ab initio} calculations of mid-mass open-shell nuclei.

\end{abstract}

\section*{PROGRAM SUMMARY}
\begin{description}[font=\normalfont\itshape]
  \item[Program title:] \texttt{ADG}
  \item[Licensing provisions:] GNU General Public License Version 3 or later
  \item[Programming language:] Python 3
  \item[Repository and DOI:] \href{https://github.com/adgproject/adg}{\nolinkurl{github.com/adgproject/adg}} \\ 
  DOI: \href{https://doi.org/10.5281/zenodo.4541534}{\nolinkurl{10.5281/zenodo.4541534}}
  \item[Nature of problem:] As formal and numerical developments in many-body-perturbation-theory-based \emph{ab initio} methods make higher orders reachable, manually producing and evaluating all the diagrams becomes rapidly untractable as both their number and complexity grow quickly. Thus, derivations are prone to mistakes and oversights.
  \item[Solution method:] BIMSRG diagrams are encoded as square matrices known as oriented adjacency matrices in graph theory, and then turned into graph objects using the \emph{NetworkX} package. These objects are used to eventually evaluate many-body expressions on a purely diagrammatic basis. The new capabilities add to those available in version (v2.0.0) of the code, i.e. the production and evaluation of HF-MBPT, diagonal and off-diagonal BMBPT diagrams.
\end{description}

\section{Introduction}
\label{sec:intro}

The intrinsic cost to solve the many-body Schr\"odinger equation scales exponentially with the particle number A. This poses a great challenge to push \textit{ab initio} calculations based on essentially exact methods such as configuration interaction~\cite{Navr09NCSMdev,Roth11SRG,Barr13PPNP} (CI) or quantum Monte-Carlo (QMC) approaches~\cite{Geze13QMCchi,Carl15RMP,Lynn17QMClight} beyond the lightest nuclei. During the last two decades the use of methods that systematically expand the exact solution with respect to a simple yet sufficiently rich A-body reference state has emerged as an invaluable paradigm to extend nuclear many-body theory to heavier masses. Indeed, truncating the expansion provides a solution whose numerical cost scales polynomially with A. The most prominent examples of such methods are many-body perturbation theory (MBPT)~\cite{Roth10PadePT,Langhammer2012,Holt14Ca,Tich16HFMBPT,Tichai:2018ncsmpt,Tichai2020review}, self-consistent Green's function (SCGF) theory~\cite{Dick04PPNP,Barb09SCGF,Cipo13Ox,Raimondi2019em,Soma2020}, coupled-cluster (CC) theory~\cite{Hagen2010cc,Hage14RPP,Bind14CCheavy,Novario2020a} and the in-medium similarity renormalization group (IMSRG) approach~\cite{Tsuk11IMSRG,Tsuk12SM,Herg13IMSRG,Bogn14SM,Herg16PR,Stro17ENO,Stroberg2019}.

Expansion methods, whenever based on a symmetry-conserving Slater-determinant reference state, efficiently describe weakly correlated closed-shell nuclei. Contrarily, they fail to describe strongly-correlated open-shell nuclei due to the inherent degeneracies of elementary excitations out of such a Slater-determinant reference state. Within the last ten years, extending these methods to symmetry-breaking reference states has shown to be a powerful idea to systematically overcome the limitation to closed-shell systems. In singly open-shell nuclei, U(1) global-gauge symmetry associated with particle-number conservation is relaxed through the use of a Bogoliubov quasiparticle vacuum as the reference state. In doubly open-shell nuclei, SU(2) rotational symmetry associated with angular-momentum conservation is (further) allowed to break via the use of deformed Slater determinants or Bogoliubov vacua as reference states. The breaking of U(1) symmetry, specifically, entails an explicit extension of the formalism based on the more general Bogoliubov algebra. This has been successfully achieved in Gorkov's extension of SCGF theory (GGF)~\cite{Soma11GGFform,Soma13GGF2N,Soma14GGF2N3N}, in Bogoliubov coupled-cluster (BCC) theory~\cite{Sign14BogCC,Henderson2014bcc} and in Bogoliubov many-body perturbation theory (BMBPT)~\cite{Duguet2015u1,Tichai18BMBPT,Arthuis2018adg1,Demol20BMBPT,Tichai2020review}. The present work similarly generalizes the single-reference IMSRG method to \emph{Bogoliubov in-medium similarity renormalization group} (BIMSRG) theory.

In recent years, various flavors of the IMSRG paradigm have led to unprecedented \textit{ab initio} calculations of atomic nuclei up to one hundred interacting particles~\cite{Morr17Tin}, of global benchmarks~\cite{Holt19atomicnucl} or of deformed nuclei~\cite{Bally2020imsrggcm}. Calculations of open-shell nuclei have been performed via the multi-reference IMSRG (MR-IMSRG)  formulated with respect to a correlated vacuum, thus requiring a more elaborate formalism than for the standard, i.e. single-reference, flavors~\cite{Herg13MR,Herg14MR,Gebr17IMNCSM}. The presently developed BIMSRG method provides a simple alternative whose advantages and drawbacks compared to MR-IMSRG can be gauged in future calculations. 

The development of novel many-body theories applicable to a large set of nuclei and the need for accurate implementations, i.e., for advanced truncation schemes, takes the derivation of the working equations to the edge of what is humanly doable. The design of BIMSRG at the BIMSRG(3) truncation level, for instance, is a perfect example with a total of 82 new diagrams to derive even when exploiting symmetries to reduce their number. Consequently, the general discussion of the BIMSRG formalism is accompanied here by the use of a new, extended version of the \textbf{\texttt{ADG}} code~\cite{Arthuis2018adg1,Arthuis2020adg2} that is capable of performing the sophisticated formal derivations in an automated fashion. 

The present work is structured as follows. The general BIMSRG formalism is introduced in Sec.~\ref{sec:qpalegbra} before being specialized to the use of normal-ordered operators with respect to a Bogoliubov vacuum in Sec.~\ref{sec:bimsrg}. Section~\ref{sec:diagrams} provides the  diagrammatic formulation of the BIMSRG that is used to design the automated derivation of the working equations at arbitrary BIMSRG(n) truncation levels through the \textbf{\texttt{ADG}} code described in Sec.~\ref{ADGcode}. The conclusions and outlook are given in Sec.~\ref{sec:outlook}, and an appendix provides the details for the BIMSRG(2) approximation.

\section{The BIMSRG formalism}
\label{sec:qpalegbra}

\subsection{Grand potential}

The second-quantized form of the Hamiltonian in an arbitrary basis $\{ c^\dagger_p, c_p \}$ of the one-body Hilbert space ${\cal H}_1$ is given by
\begin{align}
    H \equiv 
    &\phantom{+}\frac{1}{(1!)^2} \sum_{pq} \opno{t}{}{pq} c^\dagger_p c_q \notag \\
    &+\frac{1}{(2!)^2} \sum_{pqrs} \opno{v}{}{pqrs} c^\dagger_p c^\dagger_q c_s c_r \notag \\
    &+\frac{1}{(3!)^2} \sum_{pqrstu} \opno{w}{}{pqrstu} c^\dagger_p c^\dagger_q c^\dagger_r c_u c_t c_s \, , \label{H}
\end{align}
where $\opno{t}{}{pq}$ denotes matrix elements of the kinetic energy whereas $\opno{v}{}{pqrs}$ and $\opno{w}{}{pqrstu}$ denote anti-symmetric matrix elements of two- and three-nucleon interactions, respectively. The generalization to Hamiltonians with even higher many-body, e.g. four-body, interactions is straightforward (but tedious). 

In the following a constraint term involving the particle-number operator
\begin{align}
    A \equiv \sum_{p} c^\dagger_{p} c_{p}\, ,
\end{align}
is required to control the average particle-number in the many-body states of interest, hence we work with the grand potential\footnote{In practice two separate Lagrange multipliers $\mu_{\text{N}}$ and $\mu_{\text{Z}}$ are introduced to account for neutron and proton chemical potentials, such that neutron and proton number are conserved individually.},
\begin{align}
    \Omega \equiv H -\mu A \, ,
\end{align}
instead of the Hamiltonian.

\subsection{Bogoliubov algebra}

Bogoliubov quasi-particle operators are defined through a linear transformation of particle creation and annihilation operators~\cite{RingSchuck}
\begin{subequations}
\label{eq:qpbasis}
\begin{align}
\qpcre{p} &\equiv \sum_k U_{pk} c^\dagger_{k} + V_{pk} c_{k} \, , \\
\qpan{p} &\equiv \sum_k U^*_{pk} c_{k} + V_{pk}^* c^\dagger_{k} \, .
\end{align}%
\label{eq:bogotrafo}%
\end{subequations}
They obey standard Fermionic anti-commutation relations
\begin{subequations}
\begin{align}
 \{ \qpcre{p} , \qpcre{q} \} &= 0 \, , \\
 \{ \qpan{p} , \qpan{q} \} &= 0 \, , \\
 \{ \qpan{p} , \qpcre{q} \} &= \delta_{pq} \, .
\end{align}
\end{subequations}
The set of transformation coefficients $\{U,V\}$ in Eq.~\eqref{eq:qpbasis} defines a  unitary Bogoliubov transformation satisfying
\begin{subequations}
\begin{align}
U^\dagger U + V^\dagger V &= 1 \, , \\
UU^\dagger + V^\star V^T &= 1 \, , \\
U^T V + V^T U &= 0 \, , \\
U V^\dagger + V^\star U^T &= 0 \, .
\end{align}
\end{subequations}
The set of quasi-particle operators introduced in Eq.~\eqref{eq:qpbasis} also define a many-body Bogoliubov vacuum $|\Phi\ra$ through 
\begin{align}
\beta_p | \Phi \ra =0, \quad \forall p \, .
\end{align}
Due to the mixing of single-particle creation and annihilation operators in Eq.~\eqref{eq:qpbasis}, $|\Phi\ra$  breaks U(1) global-gauge symmetry associated with the conservation of particle number, i.e., it is not an eigenstate of the particle-number operator. In practice, transformation coefficients $\{U,V\}$ are typically obtained by solving Hartree-Fock-Bogoliubov (HFB) mean-field equations. 

Using $|\Phi\ra$ as the reference state,  excitations are systematically obtained via the action of an arbitrary number\footnote{The number of quasi-particle excitations is restricted to be even due to the conservation of number parity in the present work. Such a constraint should be relaxed when, e.g., building states associated with a system containing an odd number of particle via the action of an excitation operator acting on the state describing an even neighbor.} of quasi-particle creation operators
\begin{subequations}
\begin{align}
    | \Phi^{pq} \ra &\equiv \beta^\dagger_p \beta^\dagger_q | \Phi \ra \, , \\
    | \Phi^{pqrs} \ra &\equiv \beta^\dagger_p \beta^\dagger_q \beta^\dagger_r \beta^\dagger_s | \Phi \ra \, , \\
    \phantom{0} &\vdots \phantom{0}\notag \, 
\end{align}%
\label{eq:qpexc}%
\end{subequations}
 Combining the reference state with the complete set of its elementary excitations provides a basis of Fock space ${\cal F}$\footnote{Whenever the reference state reduces to a Slater determinant, the many-body states introduced in Eq.~\eqref{eq:qpexc} themselves reduce to particle-number conserving Slater determinants obtained via arbitrary numbers of n-particle/p-hole excitations of the reference state. The associated basis of Fock space becomes a direct sum of bases of Hilbert spaces ${\cal H}_N$ associated with definite particle numbers.}.

\subsection{Flow equation}

Just like standard (i.e., symmetry-conserving) single-reference IMSRG, BIMSRG is based on a unitary transformation $U(s)$ of the grand potential $\Omega$
\begin{align}
\Omega(s) \equiv U(s) \Omega U^\dagger(s) \, ,
\label{eq:simtrans}
\end{align}
that is parameterized by the continuous variable $s \in \mathbb{R}$. The similarity transformation (Eq.~\eqref{eq:simtrans}) can be re-cast into a first-order ordinary differential equation (ODE)
\begin{align}
\dds{} \Omega(s) = [\eta(s), \Omega(s)] \, ,
\label{eq:ode}
\end{align}
involving an anti-Hermitian \emph{generator} $\eta(s)$ that indirectly parametrizes the transformation. The ODE on the similarity transformed grand potential is to be solved with the initial condition $\Omega(0)=\Omega$.

The generator is chosen\footnote{The specific generators allowing for such a decoupling are not detailed in the body of the text. However, such a discussion is provided in~\ref{BIMSRG2approx} for the BIMSRG(2) approximation.} such that the reference state is decoupled from its quasi-particle excitations at the end of the flow
\begin{subequations}
\begin{align}
   \lim_{s \rightarrow \infty}  \la \Phi^{pq} | \Omega(s) | \Phi \ra &= 0 \, , \\
   \lim_{s \rightarrow \infty}  \la \Phi^{pqrs} | \Omega(s) | \Phi \ra &= 0 \, . \\
    &\vdots \notag 
\end{align}
\label{eq:decoupling}%
\end{subequations}
The exact A-body ground-state energy $\text{E}_{0}$ can then be obtained from the vacuum expectation value of the flowing grand potential
\begin{align}
\lim_{s \rightarrow \infty} \la \Phi | \Omega(s) | \Phi \ra &= \text{E}_0 -\mu \text{A} \, .
\label{eq:e0imsrg}
\end{align}

\subsection{Magnus formulation}
A different way to implement IMSRG methods is to solve for the unitary transformation itself. In the so-called Magnus formulation of the approach~\cite{Morr15Magnus}, the transformation is parameterized as
\begin{align}
    U(s) \equiv e^{\magnus(s)} \, ,
\label{eq:magnus}
\end{align}
where the Magnus operator\footnote{The present article departs from the conventional notation for the Magnus operator~\cite{Magn54exp,Blan09MagnusPR,Morr15Magnus} to avoid confusion with the grand-canonical potential introduced above.} $\magnus(s)$ satisfies $\magnus^\dagger(s) = -\magnus(s)$ and $\magnus(0) = 0$. One can now solve for $\magnus(s)$ by integrating the ODE~\cite{Magn54exp,Blan09MagnusPR}
\begin{align}
    \dds{} \magnus(s) = \sum_{l=0}^\infty  \frac{B_l}{l!} \text{ad}^{(l)}_\magnus(\eta) \, . \label{odeMagnus}
\end{align}
Here, the $B_l$ are Bernoulli numbers and $\text{ad}_\magnus(\eta)$ characterizes the adjoint action defined recursively through
\begin{subequations}
\label{adjoint}
\begin{align}
    \text{ad}^{(0)}_\magnus(\eta) &\equiv \eta(s) \, , \\
    \text{ad}^{(l)}_\magnus(\eta) &\equiv [ \magnus(s), \text{ad}^{(l-1)}_\magnus(\eta)] \, .
\end{align}
\end{subequations}

Once the Magnus operator is obtained, any arbitrary operator $O$, including $\Omega$, can be transformed consistently by using the Baker-Campbell-Hausdorff (BCH) formula
\begin{align}
    O(s) &\equiv e^{\magnus(s)} O e^{-\magnus(s)} = \sum_{k=0}^\infty \frac{1}{k!} \text{ad}_\magnus^{(k)}(O) \, . \label{BCH}
\end{align}

\subsection{Particle-number adjustment}

Because U(1) is a symmetry of the nuclear many-body Hamiltonian, i.e.,
\begin{align}
    [\Omega, A] = 0 \, ,
\end{align}
the identity
\begin{align}
    [\Omega(s), A(s)] = 0 \, \label{commut_A_flow}
\end{align}
is satisfied throughout the BIMSRG flow,\footnote{The unitary transformation itself does not fulfill the symmetry such that the commutator with the unevolved particle-number operator does not vanish for $s>0$, i.e., $[\Omega(s), A(0)] \neq 0$.} up to truncations, such that
\begin{align}
\lim_{s \rightarrow \infty} \la \Phi | A(s) | \Phi \ra &=  \text{A} \, .
\label{eq:a0imsrg}
\end{align}

Two difficulties arise to ensure that Eq.~\eqref{eq:a0imsrg} is indeed fulfilled. First, for the targeted particle number $\text{A}$ to be obtained in an \emph{exact} calculation, the chemical potential $\mu$ must be fixed such that $\text{E}_0 -\mu \text{A}$ is the lowest of all eigenvalues of $\Omega$ over Fock space. This is achievable only if the ground-state energy is strictly convex as a function of the particle number in the neighborhood of $\text{A}$, which is generally but not always true for atomic nuclei. Second, approximations made during the flow are such that the symmetry properties embodied by Eqs.~\eqref{commut_A_flow} and~\eqref{eq:a0imsrg} are explicitly violated in practice, i.e., they could only be ensured by an exact BIMSRG calculation. 

These two features require a control of Eq.~\eqref{eq:a0imsrg} throughout the flow. Following the strategies used in other expansion methods based on Bogoliubov reference states~\cite{Sign14BogCC,Tichai18BMBPT,Demol20BMBPT}, it can typically be achieved by an adjustment of the average particle number of the reference state along the flow. Providing the reference state with an explicit $s$ dependence, one allows its average particle number to be adapted along the flow 
\begin{align}
  \la \Phi(s) | A | \Phi(s) \ra =   A_\text{\text{aux}}(s) \, ,
\end{align}
such that Eq.~\eqref{eq:a0imsrg} is eventually fulfilled. Numerically, this can be achieved by iteratively solving constrained HFB calculations and BIMSRG flow equations, where the constrained value for the reference-state particle number is estimated from the BIMSRG correction until convergence is achieved.

\section{Bogoliubov normal ordering}
\label{sec:bimsrg}

Because BIMSRG relies on the use of Bogoliubov product states, algebraic equations are conveniently obtained by formulating the approach in terms of normal-ordered operators obtained via the use of Wick's theorem~\cite{Wick50theorem} with respect to the Bogoliubov vacuum $|\Phi \ra$. 

\subsection{Normal-ordered operators}

Let us consider a generic particle-number-conserving operator containing up to $N$-body terms
\begin{align}
  \label{eq:genericodef}
  O &\equiv  \sum_{n=0}^N o^{[2n]} \, ,
\end{align}
where the class $o^{[2n]}$ contains a single $n$-body operator
\begin{align}
  \label{eq:nbodydef}
  o^{[2n]}
  &\equiv o^{nn}\nonumber \\
  & = \frac{1}{n!n!} \sum_{l_1\ldots l_{2n}} o^{nn}_{l_1 \ldots l_{n} l_{n+1} \ldots l_{2n}} c^\dagger_{l_1} \ldots c^\dagger_{l_n} c_{l_{2n}} \ldots c_{l_{n+1}}  \, .
\end{align}
In Eq.~\eqref{eq:nbodydef}, $n$-body matrix elements $o^{nn}_{l_1 \ldots l_{n} l_{n+1} \ldots l_{2n}}$ are mode-$2n$ tensors, i.e., they are data arrays carrying $2n$ indices associated with the $n$ particle creation and $n$ annihilation operators they multiply. They are fully anti-symmetric with respect to the permutation of the $n$ first, resp. $n$ last, indices
\begin{equation}
  o^{nn}_{l_1 \ldots l_{n} l_{n+1} \ldots l_{2n}} = \epsilon(\sigma) \, o^{nn}_{\sigma(l_1 \ldots l_n | l_{n+1} \ldots l_{2n})} \, ,
\end{equation}
where $\epsilon(\sigma)$ refers to the signature of the permutation $\sigma$. The notation $\sigma(\ldots | \ldots)$ denotes a separation between the $n$ first and the $n$ last indices such that permutations are only considered between members of the same group. If $O$ is Hermitian, n-body matrix elements $o^{nn}_{l_1 \ldots l_n l_{n+1} \ldots l_{2n}}$ satisfy
\begin{align}
  o^{nn \ast}_{l_1 \ldots l_n l_{n+1} \ldots l_{2n}} &= o^{nn}_{l_{n+1} \ldots l_{2n} l_1 \ldots l_n} \, .
\end{align}

After normal ordering with respect to $|\Phi\rangle$, the operator $O$ can be expressed as
\begin{align}
  O
  &\equiv  \sum_{n=0}^N O^{[2n]} \, ,
\end{align}
where the class $O^{[2n]}$ groups all terms containing a normal-ordered product of $2n$ quasiparticle operators. We further differentiate these terms according to the number of quasiparticle creation and annihilation operators they contained, i.e.,
\begin{align}
  O^{[2n]}
  &\equiv \sum_{\substack{i,j=0\\i+j=2n}}^{2N} O^{ij} \, ,
\end{align}
where $O^{ij}$ gathers terms with $i$, resp. $j$, quasiparticle creation, resp. annihilation, operators and reads
\begin{align}
  O^{ij}
  &\equiv \frac{1}{i!j!} \sum_{k_1 \ldots k_{i+j}} O^{ij}_{k_1 \ldots k_{i} k_{i+1} \ldots k_{i+j}}  \, \beta^{\dagger}_{k_1} \ldots \beta^{\dagger}_{k_i} \beta_{k_{i+j}} \ldots \beta_{k_{i+1}} \, . \label{defijterm}
\end{align}
Matrix elements $O^{ij}_{k_1 \ldots k_{i+j}}$ are mode-$(i+j)$ tensors, i.e., they are data arrays carrying $i+j$ indices associated with the $i$ ($j$) quasi-particle creation (annihilation) operators they multiply. They are fully anti-symmetric with respect to the permutation of the $i$ first, resp. $j$ last, indices
\begin{align}
  O^{ij}_{k_1 \ldots k_{i} k_{i+1} \ldots k_{i+j}}
  &= \epsilon(\sigma) \, O^{ij}_{\sigma(k_1 \ldots k_i | k_{i+1} \ldots k_{i+j})} \, ,
\end{align}
and satisfy
\begin{align}
  O^{ij \ast}_{k_1 \ldots k_i k_{i+1} \ldots k_{i+j}} &= O^{ji}_{k_{i+1} \ldots k_{i+j} k_1 \ldots k_i} \, ,
\end{align}
if $O$ is Hermitian. Explicit expressions of such anti-symmetrized matrix elements up to the class $O^{[6]}$ can be found in Refs.~\cite{Sign14BogCC,Ripoche2020}.

All operators at play in the present context, e.g., $\Omega(s)$, $\eta(s)$ and $\magnus(s)$, can now be written in such a normal-ordered form. To give one explicit example, the original grand potential originating from the Hamiltonian introduced in Eq.~\eqref{H} reads
\begin{align}
    \Omega & \equiv \Omega(s=0)\notag\\
    &= \Omega^{[0]} + \Omega^{[2]} + \Omega^{[4]}+ \Omega^{[6]} \, \nonumber \\
    &\equiv \Omega^{00} +
    \opno{\Omega}{20}{} + \opno{\Omega}{11}{} + \opno{\Omega}{02}{} \notag \\
    &\phantom{=} + \opno{\Omega}{40}{} + \opno{\Omega}{31}{} + \opno{\Omega}{22}{} + \opno{\Omega}{13}{} + \opno{\Omega}{04}{} \notag \\
    &\phantom{=} + \opno{\Omega}{60}{} + \opno{\Omega}{51}{} + \opno{\Omega}{42}{} + \opno{\Omega}{33}{} + \opno{\Omega}{24}{} + \opno{\Omega}{15}{} + \opno{\Omega}{06}{} \, ,
    \label{eq:omegano}
\end{align}
i.e., it contains up to $\Omega^{[6]}$ terms.  By virtue of the normal ordering, one trivially finds that
\begin{align}
    \frac{\la \Phi | \Omega | \Phi \ra}{\la \Phi | \Phi \ra} = \opno{\Omega}{00}{}  \, .
\end{align}
Whenever $| \Phi \ra$ corresponds to the HFB vacuum, it is not coupled via $\Omega$ to two-quasi-particle excitations by virtue of Brillouin's theorem, i.e.,
\begin{align}
    \opno{\Omega}{20}{k_1k_2} = \opno{\Omega}{02}{k_1k_2} = 0, \quad \forall (k_1,k_2) \,. 
\end{align}

\subsection{Flow equation}

The BIMSRG flow equation can now be written as a system of coupled ODEs for the $ij$-components of the derivative of the grand potential, i.e.,
\begin{align}
\left(\frac{d\Omega}{ds}\right)^{ij}(s) \equiv \opno{[ \eta(s), \Omega(s) ]}{ij}{} \, , \label{ODEnormalordered}
\end{align}
where $\opno{[\,, \,]}{ij}{}$ denotes the $ij$-component of the normal-ordered operator \emph{resulting} from the commutator. Since the Hermitian adjoints of $\Omega(s)$ and $\eta(s)$ satisfy
\begin{subequations}
\begin{align}
(\opno{\Omega}{ij}{}(s))^\dagger  &= \opno{\Omega}{ji}{}(s) \, \label{eq:symmomega},\\
(\opno{\eta}{ij}{}(s))^\dagger  &= -\opno{\eta}{ji}{}(s) \, ,
\label{eq:etaantiherm}
\end{align}
\end{subequations}
the derivative of $\Omega$ is itself Hermitian, as expected. Consequently, the system of ODEs contains redundant equations and one only needs to explicitly solve for $i\geq j$ before deducing the results for $i < j$.

\subsection{Magnus formulation}

Similarly, the ODE for the Magnus operator is transformed into a set of coupled ODEs
\begin{align}
 \left(\frac{d\magnus}{ds}\right)^{ij}(s)  \equiv \sum_{l=0}^\infty  \frac{B_l}{l!} (\text{ad}^{(l)}_\magnus(\eta))^{ij} \, , \label{odeMagnusNO}
\end{align}
where $(\text{ad}^{(l)}_\magnus(\eta))^{ij}$ denotes the $ij$-component of the normal-ordered operator \emph{resulting} from the $l$-fold nested commutators. Eventually, the $ij$-component of the normal-ordered transformed operator associated with any observable $O$ of interest is obtained through
\begin{align}
    O^{ij}(s) &= \sum_{k=0}^\infty \frac{1}{k!} (\text{ad}_\magnus^{(k)}(O))^{ij} \, . \label{BCHNO}
\end{align}
As before, the (anti-)Hermiticity of $O(s)$ ($\magnus(s)$) is used to avoid redundancies.

\subsection{Approximation schemes}

The BIMSRG formalism discussed so far defines a way to exactly solve the A-body Schr\"odinger equation to access nuclear ground-states observables. As such, it is computationally intractable. For actual applications, appropriate truncation schemes have to be devised to make calculations feasible. 

The ODE in Eq.~\eqref{eq:ode} is characterized by the commutator $[ \eta(s), \Omega(s) ]$ appearing on the right-hand side. Given two operators $\opA$ and $\opB$ containing terms \emph{up to} class $\opA^{[2N_\opA]}$ and $\opB^{[2N_\opB]}$, respectively, the commutator $\opC \equiv  [\opA,\opB]$ contains terms up to class  $\opC^{[2N_\opC]}$ with $N_\opC=N_\opA + N_\opB-1$. Consequently, the maximum class of $\Omega(s)$, and thus of $\eta(s)$, increases at each propagation step of the ODE and the system of coupled ODEs \eqref{ODEnormalordered} grows rapidly. Carrying the associated operators (or, equivalently, the coefficient tensors defining them) and solving the growing numbers of coupled ODEs becomes quickly intractable. 

Adapting the truncation scheme of standard IMSRG calculations and denoting it as BIMSRG(n), only terms $O^{[2m]}$ with $m \leq n$ in all the operators at play will be retained throughout the flow. The rationale behind this truncation can be justified via perturbation theory arguments and relies on the fact that the omitted terms are (naively) of higher order than the included ones. Of course, whenever the accuracy at hand is insufficient, one must go to the next truncation level. At present, the typical working truncation scheme employed in standard IMSRG calculations is IMSRG(2) but there is a concerted effort to move towards IMSRG(3).

A similar truncation scheme can be applied to the Magnus formulation, i.e., operators in Eqs.~\eqref{odeMagnus} and~\eqref{BCH}, including those that appear in the construction of $(\text{ad}^{(l)}_\magnus(\eta))^{ij}$, are restricted to contributions $O^{[2m]}$ with $m \leq n$. The summations on the right-hand sides of Eqs.~\eqref{odeMagnus} and~\eqref{BCH} are cut off once the size of the terms drops below a desired numerical threshold \cite{Morr15Magnus} .

\section{Diagrammatic method}
\label{sec:diagrams}

The manual derivation of the BIMSRG tensor networks based on a straightforward application of Wick's theorem is already quite tedious at the BIMSRG(2) truncation level. Given that the goal is to employ more refined truncation schemes, eventually, a systematic and less error-prone methodology for deriving the BIMSRG(n) working equations is highly desirable. To set the stage for the code that will automate this the derivation (see Sec.~\ref{ADGcode}), we  now introduce a diagrammatic method for evaluating BIMSRG commutators.

\subsection{Fundamental commutator}

The IMSRG method, whether formulated through Eq.~\eqref{ODEnormalordered} or through Eqs.~\eqref{odeMagnusNO}-\eqref{BCHNO}, entirely relies on the (repeated) computation of the elementary commutator 
\begin{align}
\opC \equiv [\opA,\opB]    =  \opA \opB - \opB \opA \, , \label{elementarycomm}
\end{align}
where $\opA$ and $\opB$ denote operator whose normal-ordered contributions $\opA^{[2n]}$ ($\opB^{[2n]}$) range from $n=0$ to $n=N_\opA$ ($n=N_\opB$). 

More specifically, the goal is to compute each normal-ordered component of the operator $\opC$
\begin{align}
\opC^{ij} \equiv [\opA,\opB]^{ij}    \, , \label{elementarycomm2}
\end{align}
whose maximum class $\opC^{2N_\opC}$ is such that $N_\opC = N_\opA + N_\opB-1$\footnote{Whenever $N_\opA$ ($N_\opB$) is equal to 1, one has $N_\opC = N_\opB$ ($N_\opC = N_\opA$); i.e. it is the only situation where the maximum class of $\opC$ is not greater than those of $\opA$ and  $\opB$.}. Thus, the objective is to work out the general case characterized by the triplet $(N_\opA, N_\opB; N_\opC)$. As for the application of present interest, the derivation of the BIMSRG(n) equations relies on restricting the procedure to $N_\opA=N_\opB=n$ and to further limiting the output terms to $\opC^{[2m]}$ with $m\leq n$, i.e., it corresponds to working out the restricted case characterized by $(N_\opA, N_\opB; N_\opC) = (n, n; n)$. With this at hand, and at the price of (repeatedly) replacing $\opA$ and $\opB$ with the operators of interest, the  BIMSRG(n) algebraic equations are obtained.

\subsection{Rationale of the approach}

In principle, the computation of Eqs.~\eqref{elementarycomm}-\eqref{elementarycomm2} is straightforward: Indeed, it boils down to the application of the standard time-independent Wick's theorem to a product of two normal-ordered operators\footnote{Rigorously speaking, one is dealing with the product of two operators that are each made out of a \emph{sum} of normal-ordered contributions.}. Furthermore, the fact that one is actually interested in the \emph{commutator} of the two operators allows one to reduce the computation to so-called \emph{connected terms}, i.e., to terms where at least one elementary contraction occurs between the two operators in the application of Wick's theorem. 

A straightforward application of Wick's theorem quickly becomes cumbersome as $N_\opA$ and $N_\opB$ grow. Furthermore, it would involve repetition since many algebraic contributions are in fact identical. By identifying the corresponding patterns, one can design a diagrammatic representation of the various normal-ordered contributions to the commutator and evaluate their algebraic expressions such that a single diagram captures all these identical contributions at once. This approach eventually allows us to design an optimal code that automatically generates all needed diagrams and algebraic expressions in a matter of seconds. This will constitute the present extension of the  \textbf{\texttt{ADG}} code~\cite{Arthuis2018adg1,Arthuis2020adg2}. The diagrammatic framework introduced below shares similarities with the effective Hamiltonian diagrams occuring in BCC theory since, beyond closed vacuum-to-vacuum diagrams, linked diagrams with external legs need to be considered~\cite{Shav09MBmethod}. 

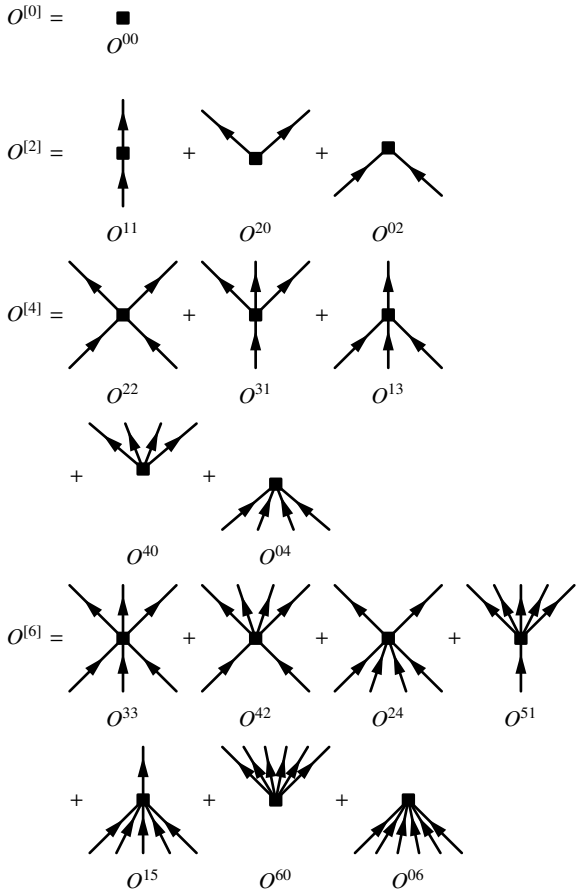
\begin{figure}[t!]
$O^{[0]} =$
	\parbox{40pt}{\begin{fmffile}{O00}
	\begin{fmfgraph*}(40,40)
	\fmftop{t1} \fmfbottom{b1}
	\fmf{phantom}{b1,i1}
	\fmf{phantom}{i1,t1}
	\fmfv{label=$O^{00}$,label.angle=-90,d.shape=square,d.filled=full,d.size=2thick}{i1}
	\end{fmfgraph*}
	\end{fmffile}}

\vspace{\baselineskip}

$O^{[2]} =$
	\parbox{40pt}{\begin{fmffile}{O11}
	\begin{fmfgraph*}(40,40)
	\fmfcmd{style_def half_prop expr p =
    draw_plain p;
    shrink(.7);
        cfill (marrow (p, .5))
    endshrink;
	enddef;}
	\fmftop{t1} \fmfbottom{b1}
	\fmf{half_prop}{b1,i1}
	\fmf{half_prop}{i1,t1}
	\fmfv{d.shape=square,d.filled=full,d.size=2thick}{i1}
	\fmflabel{$O^{11}$}{b1}
	\end{fmfgraph*}
	\end{fmffile}}
+
	\parbox{40pt}{\begin{fmffile}{O20}
	\begin{fmfgraph*}(40,40)
	\fmfcmd{style_def half_prop expr p =
    draw_plain p;
    shrink(.7);
        cfill (marrow (p, .5))
    endshrink;
	enddef;}
	\fmftop{t1,t2} \fmfbottom{b1}
	\fmfstraight
	\fmf{half_prop}{i1,t2}
	\fmf{half_prop}{i1,t1}
	\fmf{phantom,tension=2}{b1,i1}
	\fmfv{d.shape=square,d.filled=full,d.size=2thick}{i1}
	\fmflabel{$O^{20}$}{b1}
	\end{fmfgraph*}
	\end{fmffile}}
+
	\parbox{40pt}{\begin{fmffile}{O02}
	\begin{fmfgraph*}(40,40)
	\fmfcmd{style_def half_prop expr p =
    draw_plain p;
    shrink(.7);
        cfill (marrow (p, .5))
    endshrink;
	enddef;}
	\fmftop{t1} \fmfbottom{b1,b2,b3}
	\fmfstraight
	\fmf{half_prop}{b1,i1}
	\fmf{half_prop}{b3,i1}
	\fmf{phantom,tension=2}{i1,t1}
	\fmfv{d.shape=square,d.filled=full,d.size=2thick}{i1}
	\fmflabel{$O^{02}$}{b2}
	\end{fmfgraph*}
	\end{fmffile}}

\vspace{2\baselineskip}

$O^{[4]} =$
	\parbox{40pt}{\begin{fmffile}{O22}
	\begin{fmfgraph*}(40,40)
	\fmfstraight
	\fmfcmd{style_def half_prop expr p =
    draw_plain p;
    shrink(.7);
        cfill (marrow (p, .5))
    endshrink;
	enddef;}
	\fmftop{t1,t2} \fmfbottom{b1,b2,b3}
	\fmf{half_prop}{b1,i1}
	\fmf{half_prop}{b3,i1}
	\fmf{half_prop}{i1,t1}
	\fmf{half_prop}{i1,t2}
	\fmfv{d.shape=square,d.filled=full,d.size=2thick}{i1}
	\fmflabel{$O^{22}$}{b2}
	\end{fmfgraph*}
	\end{fmffile}}
+
	\parbox{40pt}{\begin{fmffile}{O31}
	\begin{fmfgraph*}(40,40)
	\fmfstraight
	\fmfcmd{style_def half_prop expr p =
    draw_plain p;
    shrink(.7);
        cfill (marrow (p, .5))
    endshrink;
	enddef;}
	\fmftop{t1,t2,t3} \fmfbottom{b1}
	\fmf{half_prop}{i1,t2}
	\fmf{half_prop}{i1,t1}
	\fmf{half_prop}{i1,t3}
	\fmf{half_prop,tension=3}{b1,i1}
	\fmfv{d.shape=square,d.filled=full,d.size=2thick}{i1}
	\fmflabel{$O^{31}$}{b1}
	\end{fmfgraph*}
	\end{fmffile}}
+
	\parbox{40pt}{\begin{fmffile}{O13}
	\begin{fmfgraph*}(40,40)
	\fmfstraight
	\fmfcmd{style_def half_prop expr p =
    draw_plain p;
    shrink(.7);
        cfill (marrow (p, .5))
    endshrink;
	enddef;}
	\fmftop{t1} \fmfbottom{b1,b2,b3}
	\fmf{half_prop}{b1,i1}
	\fmf{half_prop}{b2,i1}
	\fmf{half_prop}{b3,i1}
	\fmf{half_prop,tension=3}{i1,t1}
	\fmfv{d.shape=square,d.filled=full,d.size=2thick}{i1}
	\fmflabel{$O^{13}$}{b2}
	\end{fmfgraph*}
	\end{fmffile}}

\vspace{2\baselineskip}	

$\phantom{O^{[4]} =}$
+
	\parbox{40pt}{\begin{fmffile}{O40}
	\begin{fmfgraph*}(40,40)
	\fmfstraight
	\fmfcmd{style_def half_prop expr p =
    draw_plain p;
    shrink(.7);
        cfill (marrow (p, .5))
    endshrink;
	enddef;}
	\fmftop{t1,t2,t3,t4} \fmfbottom{b1}
	\fmf{half_prop}{i1,t1}
	\fmf{half_prop}{i1,t2}
	\fmf{half_prop}{i1,t3}
	\fmf{half_prop}{i1,t4}
	\fmf{phantom,tension=3}{i1,b1}
	\fmfv{d.shape=square,d.filled=full,d.size=2thick}{i1}
	\fmflabel{$O^{40}$}{b1}
	\end{fmfgraph*}
	\end{fmffile}}
+
	\parbox{40pt}{\begin{fmffile}{O04}
	\begin{fmfgraph*}(40,40)
	\fmfstraight
	\fmfcmd{style_def half_prop expr p =
    draw_plain p;
    shrink(.7);
        cfill (marrow (p, .5))
    endshrink;
	enddef;}
	\fmftop{t1} \fmfbottom{b1,b2,b3,b4}
	\fmf{half_prop}{b1,i1}
	\fmf{half_prop}{b2,i1}
	\fmf{half_prop}{b3,i1}
	\fmf{half_prop}{b4,i1}
	\fmf{phantom,tension=3}{i1,t1}
	\fmfv{d.shape=square,d.filled=full,d.size=2thick}{i1}
	\fmffreeze
	\fmf{phantom,label=$O^{04}$}{b2,b3}
	\end{fmfgraph*}
	\end{fmffile}}

\vspace{2\baselineskip}

$O^{[6]} =$
	\parbox{40pt}{\begin{fmffile}{O33}
	\begin{fmfgraph*}(40,40)
	\fmfstraight
	\fmfcmd{style_def half_prop expr p =
    draw_plain p;
    shrink(.7);
        cfill (marrow (p, .5))
    endshrink;
	enddef;}
	\fmftop{t1,t2,t3} \fmfbottom{b1,b2,b3}
	\fmf{half_prop}{b1,i1}
	\fmf{half_prop}{b2,i1}
	\fmf{half_prop}{b3,i1}
	\fmf{half_prop}{i1,t1}
	\fmf{half_prop}{i1,t2}
	\fmf{half_prop}{i1,t3}
	\fmfv{d.shape=square,d.filled=full,d.size=2thick}{i1}
	\fmflabel{$O^{33}$}{b2}
	\end{fmfgraph*}
	\end{fmffile}}
+
	\parbox{40pt}{\begin{fmffile}{O42}
	\begin{fmfgraph*}(40,40)
	\fmfstraight
	\fmfcmd{style_def half_prop expr p =
    draw_plain p;
    shrink(.7);
        cfill (marrow (p, .5))
    endshrink;
	enddef;}
	\fmftop{t1,t2,t3,t4} \fmfbottom{b1,b2,b3}
	\fmf{half_prop}{i1,t2}
	\fmf{half_prop}{i1,t1}
	\fmf{half_prop}{i1,t3}
	\fmf{half_prop}{i1,t4}
	\fmf{half_prop,tension=2}{b3,i1}
	\fmf{half_prop,tension=2}{b1,i1}
	\fmfv{d.shape=square,d.filled=full,d.size=2thick}{i1}
	\fmflabel{$O^{42}$}{b2}
	\end{fmfgraph*}
	\end{fmffile}}
+
	\parbox{40pt}{\begin{fmffile}{O24}
	\begin{fmfgraph*}(40,40)
	\fmfstraight
	\fmfcmd{style_def half_prop expr p =
    draw_plain p;
    shrink(.7);
        cfill (marrow (p, .5))
    endshrink;
	enddef;}
	\fmftop{t1,t2} \fmfbottom{b1,b2,b3,b4}
	\fmf{half_prop}{b1,i1}
	\fmf{half_prop}{b2,i1}
	\fmf{half_prop}{b3,i1}
	\fmf{half_prop}{b4,i1}
	\fmf{half_prop,tension=2}{i1,t2}
	\fmf{half_prop,tension=2}{i1,t1}
	\fmfv{d.shape=square,d.filled=full,d.size=2thick}{i1}
	\fmffreeze
	\fmf{phantom,label=$O^{24}$}{b2,b3}
	\end{fmfgraph*}
	\end{fmffile}}
+
	\parbox{40pt}{\begin{fmffile}{O51}
	\begin{fmfgraph*}(40,40)
	\fmfstraight
	\fmfcmd{style_def half_prop expr p =
    draw_plain p;
    shrink(.7);
        cfill (marrow (p, .5))
    endshrink;
	enddef;}
	\fmftop{t1,t2,t3,t4,t5} \fmfbottom{b1}
	\fmf{half_prop}{i1,t2}
	\fmf{half_prop}{i1,t1}
	\fmf{half_prop}{i1,t3}
	\fmf{half_prop}{i1,t4}
	\fmf{half_prop}{i1,t5}
	\fmf{half_prop,tension=5}{b1,i1}
	\fmfv{d.shape=square,d.filled=full,d.size=2thick}{i1}
	\fmflabel{$O^{51}$}{b1}
	\end{fmfgraph*}
	\end{fmffile}}
	
\vspace{2\baselineskip}	

$\phantom{O^{[6]} =}$
+
	\parbox{40pt}{\begin{fmffile}{O15}
	\begin{fmfgraph*}(40,40)
	\fmfstraight
	\fmfcmd{style_def half_prop expr p =
    draw_plain p;
    shrink(.7);
        cfill (marrow (p, .5))
    endshrink;
	enddef;}
	\fmftop{t1} \fmfbottom{b1,b2,b3,b4,b5}
	\fmf{half_prop}{b1,i1}
	\fmf{half_prop}{b2,i1}
	\fmf{half_prop}{b3,i1}
	\fmf{half_prop}{b4,i1}
	\fmf{half_prop}{b5,i1}
	\fmf{half_prop,tension=5}{i1,t1}
	\fmfv{d.shape=square,d.filled=full,d.size=2thick}{i1}
	\fmflabel{$O^{15}$}{b3}
	\end{fmfgraph*}
	\end{fmffile}}
+
	\parbox{40pt}{\begin{fmffile}{O60}
	\begin{fmfgraph*}(40,40)
	\fmfstraight
	\fmfcmd{style_def half_prop expr p =
    draw_plain p;
    shrink(.7);
        cfill (marrow (p, .5))
    endshrink;
	enddef;}
	\fmftop{t1,t2,t3,t4,t5,t6} \fmfbottom{b1}
	\fmf{half_prop}{i1,t1}
	\fmf{half_prop}{i1,t2}
	\fmf{half_prop}{i1,t3}
	\fmf{half_prop}{i1,t4}
	\fmf{half_prop}{i1,t5}
	\fmf{half_prop}{i1,t6}
	\fmf{phantom,tension=6}{i1,b1}
	\fmfv{d.shape=square,d.filled=full,d.size=2thick}{i1}
	\fmflabel{$O^{60}$}{b1}
	\end{fmfgraph*}
	\end{fmffile}}
+
	\parbox{40pt}{\begin{fmffile}{O06}
	\begin{fmfgraph*}(40,40)
	\fmfstraight
	\fmfcmd{style_def half_prop expr p =
    draw_plain p;
    shrink(.7);
        cfill (marrow (p, .5))
    endshrink;
	enddef;}
	\fmftop{t1} \fmfbottom{b1,b2,b3,b4,b5,b6}
	\fmf{half_prop}{b1,i1}
	\fmf{half_prop}{b2,i1}
	\fmf{half_prop}{b3,i1}
	\fmf{half_prop}{b4,i1}
	\fmf{half_prop}{b5,i1}
	\fmf{half_prop}{b6,i1}
	\fmf{phantom,tension=6}{i1,t1}
	\fmfv{d.shape=square,d.filled=full,d.size=2thick}{i1}
	\fmffreeze
	\fmf{phantom,label=$O^{06}$}{b3,b4}
	\end{fmfgraph*}
	\end{fmffile}}

\vspace{\baselineskip}

\caption{
\label{f:verticesO}
Diagrammatic representation of all normal-ordered contributions to an operator $O$ containing classes of terms up to $O^{[6]}$. Each term denotes a fully anti-symmetric Hugenholtz vertex. The convention is that reading an operator from left to right corresponds to an up-down reading of the associated diagram.}
\end{figure}

\subsection{Diagrammatic representation}
\label{subs:diag_represent}

The diagrammatic representation of an arbitrary operator $O$ is given by a sum of Hugenholtz vertices denoting its normal-ordered contributions $O^{ij}$. In Fig.~\ref{f:verticesO}, such a representation is displayed for an operator $O$ containing classes of terms up to $O^{[6]}$. In the following, $d_{O^{ij}} \equiv i+j$ describes the degree of a given vertex.\footnote{In the present paper, we restrict ourselves to operator of even vertex degree, as is the case when working with operators that conserve the number of particles. The formalism and code described in the following can be straightforwardly extended to incorporate operators with odd-degree vertices.} All terms $O^{ij}$ belonging to the same class $O^{[2n]}$ share the same degree $d_{O^{ij}} = 2n$. 

\begin{figure}[t!]
\begin{center}
	\parbox{40pt}{\begin{fmffile}{Avertex}
	\begin{fmfgraph*}(40,40)
	\fmftop{t1} \fmfbottom{b1}
	\fmf{phantom}{b1,i1}
	\fmf{phantom}{i1,t1}
	\fmfv{label=$\opA$,label.angle=-90,d.shape=circle,d.filled=full,d.size=3thick}{i1}
	\end{fmfgraph*}
	\end{fmffile}}
$\quad$
	\parbox{40pt}{\begin{fmffile}{Bvertex}
	\begin{fmfgraph*}(40,40)
	\fmftop{t1} \fmfbottom{b1}
	\fmf{phantom}{b1,i1}
	\fmf{phantom}{i1,t1}
	\fmfv{label=$\opB$,label.angle=-90,d.shape=circle,d.filled=empty,d.size=3thick}{i1}
	\end{fmfgraph*}
	\end{fmffile}}
$\quad$
	\parbox{40pt}{\begin{fmffile}{Cvertex}
	\begin{fmfgraph*}(40,40)
	\fmfcmd{
    		path quadrant, q[], otimes;
    		quadrant = (0, 0) -- (0.5, 0) & quartercircle & (0, 0.5) -- (0, 0);
    		for i=1 upto 4: q[i] = quadrant rotated (45 + 90*i); endfor
    		otimes = q[1] & q[2] & q[3] & q[4] -- cycle;
	}
	\fmfwizard
	\fmftop{t1} \fmfbottom{b1}
	\fmf{phantom}{b1,i1}
	\fmf{phantom}{i1,t1}
	\fmfv{label=$\opC$,label.angle=-90,d.shape=otimes,d.filled=empty,d.size=3thick}{i1}
	\end{fmfgraph*}
	\end{fmffile}}
\end{center}

\caption{
\label{f:verticesABC}
Vertices associated with operators $\opA$, $\opB$ and $\opC$.}
\end{figure}
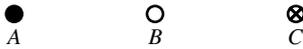

In the diagrammatic method presented below, a graphical representation of each of the three operators $\opA$, $\opB$ and $\opC$ occuring in Eq.~\eqref{elementarycomm} is needed. The corresponding vertices are introduced in Fig.~\ref{f:verticesABC}.

\begin{figure}[t!]
\begin{center}
\parbox{40pt}{\begin{fmffile}{OrientRule}
	\begin{fmfgraph*}(40,40)
	\fmfstraight
	\fmfcmd{style_def half_prop expr p =
    draw_plain p;
    shrink(.7);
        cfill (marrow (p, .5))
    endshrink;
	enddef;}
	\fmftop{t1,t2} \fmfbottom{b1,b2}
	\fmfv{label=$-O^{22}_{k_1 k_2 k_3 k_4}$,label.angle=20,label.dist=15pt}{i1}
	\fmf{half_prop}{b1,i1}
	\fmf{half_prop}{b2,i1}
	\fmf{half_prop}{i1,t1}
	\fmf{half_prop}{i1,t2}
	\fmfv{d.shape=square,d.filled=full,d.size=2thick}{i1}
	\fmfv{l=$k_2$,l.d=0.05w}{t1}
	\fmfv{l=$k_1$,l.d=0.05w}{t2}
	\fmfv{l=$k_3$,l.d=0.03w}{b1}
	\fmfv{l=$k_4$,l.d=0.05w}{b2}
	\fmffreeze
	\end{fmfgraph*}
	\end{fmffile}}
\hspace{40pt} = \hspace{5pt}
\parbox{40pt}{\begin{fmffile}{OrientRule}
	\begin{fmfgraph*}(40,40)
	\fmfstraight
	\fmfcmd{style_def half_prop expr p =
    draw_plain p;
    shrink(.7);
        cfill (marrow (p, .5))
    endshrink;
	enddef;}
	\fmftop{t1,t2} \fmfbottom{b1,b2}
	\fmfv{label=$+O^{22}_{k_1 k_2 k_3 k_4}$,label.angle=20,label.dist=15pt}{i1}
	\fmf{half_prop}{b1,i1}
	\fmf{half_prop}{b2,i1}
	\fmf{half_prop}{i1,t1}
	\fmf{half_prop}{i1,t2}
	\fmfv{d.shape=square,d.filled=full,d.size=2thick}{i1}
	\fmfv{l=$k_1$,l.d=0.05w}{t1}
	\fmfv{l=$k_2$,l.d=0.05w}{t2}
	\fmfv{l=$k_3$,l.d=0.03w}{b1}
	\fmfv{l=$k_4$,l.d=0.05w}{b2}
	\fmffreeze
	\end{fmfgraph*}
	\end{fmffile}}
\end{center}
\caption{
\label{variousvertices3}
Left: canonical labelling of the legs and associated vertex in the diagrammatic representation of the operator $O^{22}$. Right: sign rule to apply when departing from the canonical labelling through the crossing of the legs associated with the two creation operators.}
\end{figure}
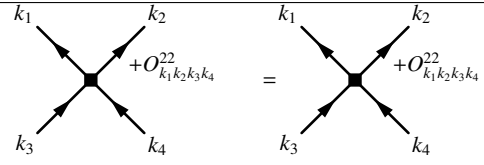

The diagrammatic representation of $O^{ij}$ involves the vertex itself along with $i$ ($j$) lines traveling out of (into) the vertex and representing quasi-particle creation (annihilation) operators. The diagram is meant to be labelled in order to eventually translate it into an algebraic expression. The canonical rule to do so consists of first assigning indices $k_1 \ldots k_i$ consecutively from the leftmost to the rightmost line above the vertex, while $k_{i+1} \ldots k_{i+j}$ must be similarly assigned consecutively to lines below the vertex. Next, the anti-symmetric matrix element $O^{ij}_{k_1 \ldots k_i k_{i+1} \ldots k_{i+j}}$ is to be assigned to the vertex. This canonical labelling is illustrated in Fig.~\ref{variousvertices3} for the diagram representing the operator $O^{22}$. 

In case the left-right labelling of the incoming and/or outgoing legs does not match the canonical ordering employed in the amplitude $O^{ij}_{k_1 \ldots k_i k_{i+1} \ldots k_{i+j}}$, the latter must be multiplied with the sign $(-1)^{l_p}$, where $l_p$ denotes the signature of the permutation of the incoming and/or outgoing legs necessary to bring them into the canonical ordering. This additional rule is illustrated in Fig.~\ref{variousvertices3} for $O^{22}$.

\begin{figure}[t!]

\vspace{1\baselineskip}

\begin{center}

$R^{-+}_{k_1k_2} = $
\parbox{40pt}{\begin{fmffile}{R_detailed}
\begin{fmfgraph*}(40,40)
\fmfcmd{style_def prop_pm expr p =
    draw_plain p;
    shrink(.7);
        cfill (marrow (p, .25));
        cfill (marrow (p, .75))
    endshrink;
enddef;
}
\fmftop{t1} \fmfbottom{b1}
\fmflabel{\small $k_1$}{t1}
\fmflabel{\small $k_2$}{b1}
\fmf{prop_pm}{b1,t1}
\end{fmfgraph*}
\end{fmffile}}
$=$
\parbox{40pt}{\begin{fmffile}{R_simple}
\begin{fmfgraph*}(40,40)
\fmfcmd{style_def half_prop expr p =
    draw_plain p;
    shrink(.7);
        cfill (marrow (p, .5))
    endshrink;
	enddef;
}
\fmftop{t1} \fmfbottom{b1}
\fmf{half_prop, label={\small $k_1$}}{b1,t1}
\end{fmfgraph*}
\end{fmffile}}

\end{center}

\caption{
\label{f:prop}
Diagrammatic representation of the single non-zero elementary contraction. The convention is that the left-to-right reading of a matrix element corresponds to the up-down reading of the diagram.}
\end{figure}
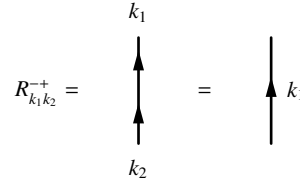

Beyond operators, the elementary contractions occurring in the standard Wick's theorem need to be represented diagrammatically. Given that the operators are conveniently expressed in the quasi-particle basis associated with the Bogoliubov vacuum, the four elementary contractions take the simplest possible form
\begin{align}
\mathbf{R}_{k_1k_2} &=
\begin{pmatrix}
\frac{\langle \Phi |\, \beta^{\dagger}_{k_1} \beta^{\phantom{\dagger}}_{k_2} | \Phi \rangle}{\langle \Phi | \Phi \rangle} & \frac{\langle \Phi |\, \beta^{\phantom{\dagger}}_{k_1} \beta^{\phantom{\dagger}}_{k_2} | \Phi \rangle}{\langle \Phi | \Phi \rangle} \\
\frac{\langle \Phi |\, \beta^{\dagger}_{k_1} \beta^{\dagger}_{k_2} | \Phi \rangle}{\langle \Phi | \Phi \rangle} &  \frac{\langle \Phi |\, \beta^{\phantom{\dagger}}_{k_1} \beta^{\dagger}_{k_2} | \Phi \rangle}{\langle \Phi | \Phi \rangle}
\end{pmatrix} \nonumber \\
&\equiv
\begin{pmatrix}
R^{+-}_{k_1k_2} & R^{--}_{k_1k_2} \\
R^{++}_{k_1k_2} &  R^{-+}_{k_1k_2}
\end{pmatrix} \nonumber \\
&=
\begin{pmatrix}
0 & 0 \\
0 &  \delta_{k_1k_2}
\end{pmatrix} \ , \label{generalizeddensitymatrix2}
\end{align}
such that the sole non-zero contraction $R^{-+}_{k_1k_2} = \delta_{k_1k_2}$ needs to be considered. The diagrammatic representation of this contraction is provided in Fig.~\ref{f:prop} and connects two up-going lines associated with one annihilation and one creation operator, both carrying the same quasi-particle index. For simplicity, one can eventually represent the contraction as a line carrying a single up-going arrow along with one quasi-particle index.

The complete set of diagrams that contribute to $\opC=[\opA,\opB]$ can now be systematically generated and evaluated by assembling the building blocks according to a specific set of rules. In principle, both contributions $+\opA\opB$ and $-\opB\opA$ need to be processed explicitly. However, given that the first term is considered for arbitrary values of $N_\opA$ and $N_\opB$, the diagrams making up the second term can easily be deduced from those contributing to the first one by (i) adding a global minus sign and (ii) performing a systematic exchange $\opA \leftrightarrow \opB$. Thus, the rules to produce and evaluate relevant diagrams are only specified for $+\opA\opB$ below while it is understood that this term is to be combined with the diagrams for $-\opB\opA$ eventually. As explained below, summing the two terms making up the commutator leads to the cancellation of a specific category of contributions. 

\subsection{Diagram generation}
\label{subs:handmande_diag_gene}

The complete set of relevant diagrams are generated from the basic building blocks by applying the following set of topological rules:
\begin{enumerate}
\item A diagram contributing to $\opC$ combines a top vertex $\opA$ and a bottom vertex $\opB$ connected by a number $n_l$ of elementary contractions $R^{-+}$. Because $\opA$ and $\opB$ are normal ordered, contractions starting and ending at the same vertex cannot be of the $R^{-+}$ type and are therefore necessarily zero according to Eq.~\eqref{generalizeddensitymatrix2}. Consequently, each contraction necessarily connects $\opA$ and $\opB$.
\item The fact that $\opC$ actually represents the \emph{commutator} of $\opA$ and $\opB$ immediately forces all relevant diagrams to be \emph{connected} \footnote{The fact that BIMSRG(n) equations arise from a commutator ensures that the associated results are \emph{size-extensive}, i.e., that truncation errors from omitting diagrams scale linearly with the system size.}, i.e., in each diagrams $\opA$ and  $\opB$ are connected by \emph{at least} one contraction and $n_l >0$. Indeed, terms with $n_l =0$ contribute identically to $\opA\opB$ and $\opB\opA$ and cancel out. Consequently, contributions involving $\opA^{00}$ and $\opB^{00}$ along with the class of terms $\opC^{[2(N_\opA+N_\opB)]}$ can be omitted from the outset. As a result,
\begin{enumerate}
\item Normal-ordered contributions $\opA^{kl}$  and  $\opB^{mn}$ can be limited to classes $d_{\opA^{kl}} \equiv k+l=2,\ldots,2N_\opA$ and $d_{\opB^{mn}} \equiv m+n=2,\ldots,2N_\opB$, respectively.
\item $\opC$ receives non-zero contributions $\opC^{ij}$ for $d_{\opC^{ij}} \equiv i+j =0,\ldots, 2(N_\opA+N_\opB-1)$.
\end{enumerate}
\item To generate all relevant contributions to $\opC^{ij}$, normal-ordered contributions to $\opA$ and $\opB$ must be contracted in all possible ways. Not all combinations of $\opA^{kl}$ and $\opB^{mn}$ can actually contribute to a given term  $\opC^{ij}$: The condition 
\begin{equation}
i-j = (k-l) + (m-n)  \, , \label{identity1}
\end{equation}
must be fulfilled, which allows one to discard a large set of combinations from the outset. Whenever Eq.~\eqref{identity1} is satisfied, the number of contractions involved in the diagram denoted by $C^{ij}(kl,mn)$ is
\begin{equation}
n_l = k+m-i = l+n-j  \, .  \label{identity2}
\end{equation}
\item Specific properties of $\opA$ and $\opB$ typically translate into properties of $\opC$ such that only a subset of the normal-ordered contributions $\opC^{ij}$ has to be computed explicitly. For example, if $\opA$ is anti-Hermitian and $\opB$ is Hermitian, $\opC$ is Hermitian and it is sufficient to only evaluate the subset of terms  $\{\opC^{ij}, i\geq j\}$.
\item A convenient way to classify diagrams is to introduce the maximal vertex degree of a diagram, defined as $d_\mathrm{max} \equiv \max(d_{\opA^{kl}}, d_{\opB^{mn}}, d_{\opC^{ij}})$. In particular, the diagrams arising for the first time at order BIMSRG(n) correspond to all the topologies characterized by $d_\mathrm{max} = n$.
\end{enumerate}

\subsection{Diagram evaluation}

Once the complete set of diagrams contributing to $\opC^{ij}$ is generated, the expressions of the associated \emph{amplitudes}, i.e., of the anti-symmetrized matrix elements $\opC^{ij}_{k_1 \ldots k_{i} k_{i+1} \ldots k_{i+j}}$, are obtained by applying a set of algebraic rules to each contributing diagram
\begin{enumerate}
\item Label outgoing (incoming) external lines with quasi-particle indices $k_1 \ldots k_i $ ($k_{i+1} \ldots k_{i+j}$).
\item Label the $n_l$ internal lines with different quasi-particle indices, e.g., $p,q,r,...$.
\item Attribute to vertices  $\opA^{kl}$ and $\opB^{mn}$ their associated canonical amplitudes.  
\item Sum over the $n_l$ internal line labels. 
\item Include a factor $(n_l!)^{-1}$ for the equivalent internal lines. Equivalent internal lines begin and end at the same two vertices. By virtue of the fact that each diagram contains only two vertices and that there is no self contraction, the $n_l$ internal lines are necessarily \emph{equivalent} in the present context.
\item Multiply by a phase factor $(-1)^{n_c}$, where $n_c$ is the number of line crossings 
in the diagram (vertices are not considered line crossings).
\item Sum over all distinct permutations $P$ of labels of inequivalent incoming (outgoing) external lines, 
including a parity factor $(-1)^{\sigma(P)}$ associated with the signature $\sigma(P)$ of the permutation. Incoming (outgoing) external lines are inequivalent if they do \emph{not} originate from the same vertex.
\end{enumerate}
These algebraic rules are fully consistent with the ones for BCC~\cite{Sign14BogCC} since they originate from the application of the standard Wick's theorem under the same conditions\footnote{In BCC, there exists one more rule associated with so-called equivalent cluster amplitudes.}.

\subsection{Elementary example}

\begin{figure}[t!]
\vspace{\baselineskip}
\begin{center}
\parbox{80pt}{\begin{fmffile}{Example}
\begin{fmfgraph*}(50,80)
\fmfcmd{style_def half_prop expr p =
    draw_plain p;
    shrink(.7);
        cfill (marrow (p, .5))
    endshrink;
	enddef;
}
\fmfstraight
\fmftopn{t}{4} \fmfbottom{b1}
\fmflabel{\small $k_1$}{t2}
\fmflabel{\small $k_2$}{t3}
\fmfv{label=\small $k_3$,l.angle=90}{t4}
\fmflabel{\small $k_4$}{b1}
\fmfv{label=\opA$^{(22)}_{k_1k_2 p q}$,label.angle=180,d.shape=circle,d.filled=full,d.size=3thick}{i2}
\fmfv{label=\opB$^{(31)}_{pq k_3k_4}$,label.angle=180,d.shape=circle,d.filled=empty,d.size=3thick}{i1}
\fmf{half_prop}{b1,i1}
\fmf{half_prop,tension=0.5}{i2,t2}
\fmf{half_prop,tension=0.5}{i2,t3}
\fmf{half_prop,tension=0.3,left=0.5,label=\small $p$, l.dist=4pt}{i1,i2}
\fmf{half_prop,tension=0.3,right=0.5,label=\small $q$, l.dist=4pt,l.side=left}{i1,i2}
\fmffreeze
\fmf{half_prop,right=0.25}{i1,t4}
\end{fmfgraph*}
\end{fmffile}}
\end{center}

\caption{
\label{f:example}
Fully-labeled form of diagram $\opC^{31}(22,31)$.}
\end{figure}
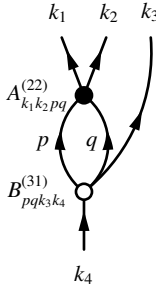

Let us demonstrate the use of both the topological and algebraic rules for a simple example. For this purpose, we consider the diagram built out of $\opA^{(22)}$ and $\opB^{(31)}$ and contributing to $\opC^{31}$, i.e., diagram $\opC^{31}(22,31)$. Such a combination is indeed allowed by Eq.~\eqref{identity1} and the corresponding fully labeled diagram is displayed in Fig.~\ref{f:example}. Contracting $\opA^{(22)}$ and $\opB^{(31)}$ consistently with Eq.~\eqref{identity2}, $n_l=2$ internal lines are necessary to obtain three outgoing external lines, and one incoming.

The diagram possesses one pair $(p,q)$ of equivalent internal lines and two pairs $(k_1,k_3)$ and $(k_2,k_3)$ of inequivalent external lines. As drawn, the amplitudes are given in canonical form and the diagram displays no line crossing. As a result, the algebraic expression of the anti-symmetrized matrix element is given by
\begin{align}
\opC^{31}_{k_1k_2k_3k_4}(22,31) =    \permtwo{k_1k_2}{k_3} \frac{1}{2} \sum_{pq} 
   \opA^{22}_{k_1k_2 pq} \opB^{31}_{pq k_3k_4}
    \, . \label{expressionexample}
\end{align}

In Eq.~\eqref{expressionexample}, the permutation operator $\permtwo{s_1}{s_2}=\permtwo{s_2}{s_1}$ generically performs an appropriate anti-symmetrization of the matrix element it acts on. The operator does so by permuting all indices in set $s_1$ with all indices in set $s_2$ while including a sign associated with the signature of the permutation. In the above example, the operator is defined by
\begin{align}
    \permtwo{k_1k_2}{k_3} &\equiv 1 - P_{k_1k_3} - P_{k_2 k_3}\, ,
\end{align}
where, e.g., $P_{k_1k_3}$ commutes indices $k_1$ and $k_3$. Because the elementary commutator involves two operators, the required permutation operators necessarily involve only \emph{two} sets of indices; i.e. $s_1$ and $s_2$. This is at variance with, e.g., BCC where a larger number of operators may appear such that permutation operators involving more than two sets of indices are called for~\cite{Sign14BogCC}.\footnote{In, e.g., BCC, contributions may involve more than two operators with external legs, e.g., double-excitation amplitudes constructed from one $\mathcal{T}_1$, one $\mathcal{T}_2$ operator and a two-body vertex of the grand-canonical potential, as the diagram labelled D6c in Ref.~\cite{Sign14BogCC}. As such, permutation operators between more than two sets of indices appear in such formalisms.} Still, the \emph{content} of the two sets of indices presently depend on the classes of terms entering $\opA$, $\opB$ and $\opC$. As an example, the complete list of permutation operators involved at the BIMSRG(2) level is provided in Eq.~\eqref{permutoperators}.

Because $\opC^{31}_{k_1k_2k_3k_4}(22,31)$ is anti-symmetric with respect to the exchange of any pair of labels within the triplet $(k_1,k_2,k_3)$, it is sufficient to compute it explicitly for, e.g., the subset $k_1<k_2<k_3$.

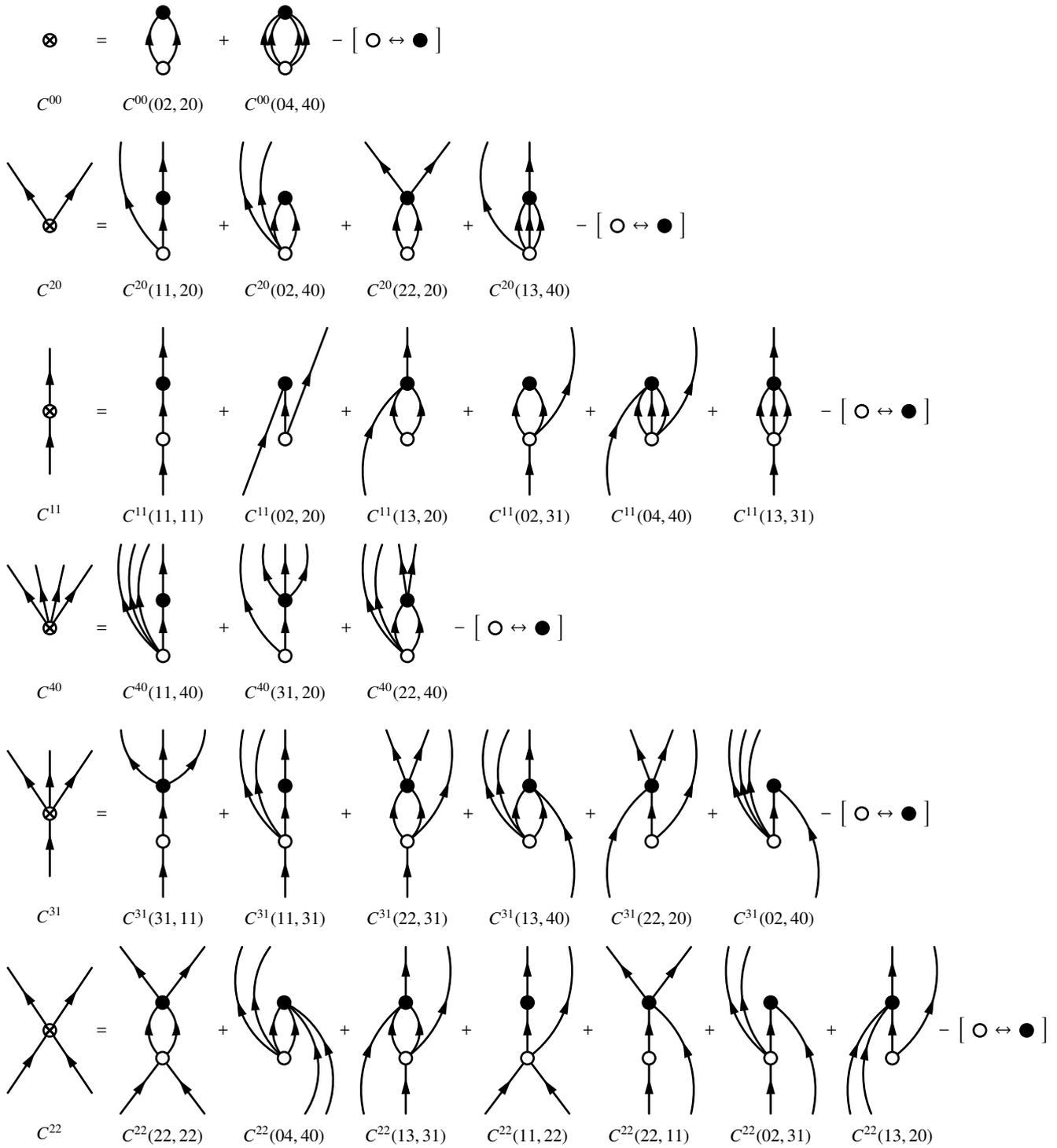
\begin{figure*}[t!]
\vspace{-15pt}
\parbox{40pt}{\begin{fmffile}{C00}
\begin{fmfgraph*}(40,60)
\fmfcmd{style_def half_prop expr p =
    draw_plain p;
    shrink(.7);
        cfill (marrow (p, .5))
    endshrink;
	enddef;
}
\fmfcmd{
    	path quadrant, q[], otimes;
    	quadrant = (0, 0) -- (0.5, 0) & quartercircle & (0, 0.5) -- (0, 0);
    	for i=1 upto 4: q[i] = quadrant rotated (45 + 90*i); endfor
    	otimes = q[1] & q[2] & q[3] & q[4] -- cycle;
}
\fmfwizard
\fmfstraight
\fmftop{t1} \fmfbottom{b1}
\fmfv{d.shape=otimes,d.filled=empty,d.size=3thick}{i1}
\fmf{phantom}{b1,i1}
\fmf{phantom}{i1,t1}
\fmfv{label=\opC$^{00}$,label.d=0pt}{b1}
\end{fmfgraph*}
\end{fmffile}}
$ = \enspace$
\parbox{40pt}{\begin{fmffile}{C00_1}
\begin{fmfgraph*}(40,80)
\fmfcmd{style_def half_prop expr p =
    draw_plain p;
    shrink(.7);
        cfill (marrow (p, .5))
    endshrink;
	enddef;
}
\fmfstraight
\fmftop{t1} \fmfbottom{b1}
\fmfv{d.shape=circle,d.filled=full,d.size=3thick}{i2}
\fmfv{d.shape=circle,d.filled=empty,d.size=3thick}{i1}
\fmf{phantom}{b1,i1}
\fmf{half_prop,left=0.5,tension=0.5}{i1,i2}
\fmf{half_prop,right=0.5,tension=0.5}{i1,i2}
\fmf{phantom}{i2,t1}
\fmfv{label=\opC$^{00}(02,,20)$,label.d=5pt,l.angle=90}{b1}
\end{fmfgraph*}
\end{fmffile}}
$\enspace + \enspace$
\parbox{40pt}{\begin{fmffile}{C00_2}
\begin{fmfgraph*}(40,80)
\fmfcmd{style_def half_prop expr p =
    draw_plain p;
    shrink(.7);
        cfill (marrow (p, .5))
    endshrink;
	enddef;
}
\fmfstraight
\fmftop{t1} \fmfbottom{b1}
\fmfv{d.shape=circle,d.filled=full,d.size=3thick}{i2}
\fmfv{d.shape=circle,d.filled=empty,d.size=3thick}{i1}
\fmf{phantom}{b1,i1}
\fmf{half_prop,left=0.5,tension=0.25}{i1,i2}
\fmf{half_prop,right=0.5,tension=0.25}{i1,i2}
\fmf{half_prop,left=0.75,tension=0.25}{i1,i2}
\fmf{half_prop,right=0.75,tension=0.25}{i1,i2}
\fmf{phantom}{i2,t1}
\fmfv{label=\opC$^{00}(04,,40)$,label.d=5pt,l.angle=90}{b1}
\end{fmfgraph*}
\end{fmffile}}
$- \enspace \Big[$
\parbox{10pt}{\begin{fmffile}{B_small}
\begin{fmfgraph*}(10,10)\fmfkeep{Bvertex}
\fmftop{t1} \fmfbottom{b1}
\fmfv{d.shape=circle,d.filled=empty,d.size=3thick}{i1}
\fmf{phantom}{b1,i1}
\fmf{phantom}{i1,t1}
\end{fmfgraph*}
\end{fmffile}}
$\leftrightarrow$
\parbox{10pt}{\begin{fmffile}{A_small}
\begin{fmfgraph*}(10,10)\fmfkeep{Avertex}
\fmftop{t1} \fmfbottom{b1}
\fmfv{d.shape=circle,d.filled=full,d.size=3thick}{i1}
\fmf{phantom}{b1,i1}
\fmf{phantom}{i1,t1}
\end{fmfgraph*}
\end{fmffile}}
$\Big]$

\vspace{.8\baselineskip}

\parbox{40pt}{\begin{fmffile}{C20}
\begin{fmfgraph*}(40,60)
\fmfcmd{style_def half_prop expr p =
    draw_plain p;
    shrink(.7);
        cfill (marrow (p, .5))
    endshrink;
	enddef;
}
\fmfcmd{
    	path quadrant, q[], otimes;
    	quadrant = (0, 0) -- (0.5, 0) & quartercircle & (0, 0.5) -- (0, 0);
    	for i=1 upto 4: q[i] = quadrant rotated (45 + 90*i); endfor
    	otimes = q[1] & q[2] & q[3] & q[4] -- cycle;
}
\fmfwizard
\fmfstraight
\fmftopn{t}{2} \fmfbottom{b1}
\fmfv{d.shape=otimes,d.filled=empty,d.size=3thick}{i1}
\fmf{phantom}{b1,i1}
\fmf{half_prop,tension=0.5}{i1,t1}
\fmf{half_prop,tension=0.5}{i1,t2}
\fmfv{label=\opC$^{20}$,label.d=0pt}{b1}
\end{fmfgraph*}
\end{fmffile}}
$ = \enspace$
\parbox{40pt}{\begin{fmffile}{C20_1}
\begin{fmfgraph*}(40,80)
\fmfcmd{style_def half_prop expr p =
    draw_plain p;
    shrink(.7);
        cfill (marrow (p, .5))
    endshrink;
	enddef;
}
\fmfstraight
\fmftopn{t}{3} \fmfbottom{b1}
\fmfv{d.shape=circle,d.filled=full,d.size=3thick}{i2}
\fmfv{d.shape=circle,d.filled=empty,d.size=3thick}{i1}
\fmf{phantom}{b1,i1}
\fmf{half_prop}{i2,t2}
\fmf{half_prop}{i1,i2}
\fmffreeze
\fmf{half_prop,left=0.25}{i1,t1}
\fmfv{label=\opC$^{20}(11,,20)$,label.d=5pt,l.angle=90}{b1}
\end{fmfgraph*}
\end{fmffile}}
$\enspace + \enspace$
\parbox{40pt}{\begin{fmffile}{C20_2}
\begin{fmfgraph*}(40,80)
\fmfcmd{style_def half_prop expr p =
    draw_plain p;
    shrink(.7);
        cfill (marrow (p, .5))
    endshrink;
	enddef;
}
\fmfstraight
\fmftopn{t}{4} \fmfbottom{b1}
\fmfv{d.shape=circle,d.filled=full,d.size=3thick}{i2}
\fmfv{d.shape=circle,d.filled=empty,d.size=3thick}{i1}
\fmf{phantom}{b1,i1}
\fmf{phantom,tension=0.5}{i2,t2}
\fmf{phantom,tension=0.5}{i2,t3}
\fmf{half_prop,left=0.4,tension=0.5}{i1,i2}
\fmf{half_prop,right=0.4,tension=0.5}{i1,i2}
\fmffreeze
\fmf{half_prop,left=0.3}{i1,t1}
\fmf{half_prop,left=0.3}{i1,t2}
\fmfv{label=\opC$^{20}(02,,40)$,label.d=5pt,l.angle=90}{b1}
\end{fmfgraph*}
\end{fmffile}}
$\enspace + \enspace$
\parbox{40pt}{\begin{fmffile}{C20_3}
\begin{fmfgraph*}(40,80)
\fmfcmd{style_def half_prop expr p =
    draw_plain p;
    shrink(.7);
        cfill (marrow (p, .5))
    endshrink;
	enddef;
}
\fmfstraight
\fmftopn{t}{2} \fmfbottom{b1}
\fmfv{d.shape=circle,d.filled=full,d.size=3thick}{i2}
\fmfv{d.shape=circle,d.filled=empty,d.size=3thick}{i1}
\fmf{phantom}{b1,i1}
\fmf{half_prop,tension=0.5}{i2,t1}
\fmf{half_prop,tension=0.5}{i2,t2}
\fmf{half_prop,left=0.4,tension=0.5}{i1,i2}
\fmf{half_prop,right=0.4,tension=0.5}{i1,i2}
\fmfv{label=\opC$^{20}(22,,20)$,label.d=5pt,l.angle=90}{b1}
\end{fmfgraph*}
\end{fmffile}}
$\enspace + \enspace $
\parbox{40pt}{\begin{fmffile}{C20_4}
\begin{fmfgraph*}(40,80)
\fmfcmd{style_def half_prop expr p =
    draw_plain p;
    shrink(.7);
        cfill (marrow (p, .5))
    endshrink;
	enddef;
}
\fmfstraight
\fmftopn{t}{3} \fmfbottom{b1}
\fmfv{d.shape=circle,d.filled=full,d.size=3thick}{i2}
\fmfv{d.shape=circle,d.filled=empty,d.size=3thick}{i1}
\fmf{phantom}{b1,i1}
\fmf{half_prop}{i1,i2}
\fmf{half_prop}{i2,t2}
\fmffreeze
\fmf{half_prop,left=0.4}{i1,i2}
\fmf{half_prop,right=0.4}{i1,i2}
\fmf{half_prop,left=0.4}{i1,t1}
\fmfv{label=\opC$^{20}(13,,40)$,label.d=5pt,l.angle=90}{b1}
\end{fmfgraph*}
\end{fmffile}}
$- \enspace \Big[$
\parbox{10pt}{\fmfreuse{Bvertex}}
$\leftrightarrow$
\parbox{10pt}{\fmfreuse{Avertex}}
$\Big]$

\vspace{.8\baselineskip}

\parbox{40pt}{\begin{fmffile}{C11}
\begin{fmfgraph*}(40,60)
\fmfcmd{style_def half_prop expr p =
    draw_plain p;
    shrink(.7);
        cfill (marrow (p, .5))
    endshrink;
	enddef;
}
\fmfcmd{
    	path quadrant, q[], otimes;
    	quadrant = (0, 0) -- (0.5, 0) & quartercircle & (0, 0.5) -- (0, 0);
    	for i=1 upto 4: q[i] = quadrant rotated (45 + 90*i); endfor
    	otimes = q[1] & q[2] & q[3] & q[4] -- cycle;
}
\fmfwizard
\fmfstraight
\fmftop{t1} \fmfbottom{b1}
\fmfv{d.shape=otimes,d.filled=empty,d.size=3thick}{i1}
\fmf{half_prop}{b1,i1}
\fmf{half_prop}{i1,t1}
\fmfv{label=\opC$^{11}$,l.dist=15pt}{b1}
\end{fmfgraph*}
\end{fmffile}}
$ = \enspace$
\parbox{40pt}{\begin{fmffile}{C11_1}
\begin{fmfgraph*}(40,80)
\fmfcmd{style_def half_prop expr p =
    draw_plain p;
    shrink(.7);
        cfill (marrow (p, .5))
    endshrink;
	enddef;
}
\fmfstraight
\fmftopn{t}{3} \fmfbottom{b1}
\fmfv{d.shape=circle,d.filled=full,d.size=3thick}{i2}
\fmfv{d.shape=circle,d.filled=empty,d.size=3thick}{i1}
\fmf{half_prop}{b1,i1}
\fmf{half_prop}{i2,t2}
\fmf{half_prop}{i1,i2}
\fmfv{label=\opC$^{11}(11,,11)$}{b1}
\end{fmfgraph*}
\end{fmffile}}
$\enspace + \enspace$
\parbox{40pt}{\begin{fmffile}{C11_2}
\begin{fmfgraph*}(40,80)
\fmfcmd{style_def half_prop expr p =
    draw_plain p;
    shrink(.7);
        cfill (marrow (p, .5))
    endshrink;
	enddef;
}
\fmfstraight
\fmftopn{t}{3} \fmfbottomn{b}{3}
\fmfv{d.shape=circle,d.filled=full,d.size=3thick}{i2}
\fmfv{d.shape=circle,d.filled=empty,d.size=3thick}{i1}
\fmf{phantom}{b2,i1}
\fmf{phantom}{i2,t2}
\fmf{half_prop}{i1,i2}
\fmffreeze
\fmf{half_prop}{b1,i2}
\fmf{half_prop}{i1,t3}
\fmfv{label=\opC$^{11}(02,,20)$}{b2}
\end{fmfgraph*}
\end{fmffile}}
$\enspace + \enspace$
\parbox{40pt}{\begin{fmffile}{C11_3}
\begin{fmfgraph*}(40,80)
\fmfcmd{style_def half_prop expr p =
    draw_plain p;
    shrink(.7);
        cfill (marrow (p, .5))
    endshrink;
	enddef;
}
\fmfstraight
\fmftop{t1} \fmfbottomn{b}{3}
\fmfv{d.shape=circle,d.filled=full,d.size=3thick}{i2}
\fmfv{d.shape=circle,d.filled=empty,d.size=3thick}{i1}
\fmf{phantom}{b2,i1}
\fmf{half_prop}{i2,t1}
\fmf{half_prop,right=0.5,tension=0.5}{i1,i2}
\fmf{half_prop,left=0.5,tension=0.5}{i1,i2}
\fmffreeze
\fmf{half_prop,left=0.3}{b1,i2}
\fmfv{label=\opC$^{11}(13,,20)$}{b2}
\end{fmfgraph*}
\end{fmffile}}
$\enspace + \enspace$
\parbox{40pt}{\begin{fmffile}{C11_4}
\begin{fmfgraph*}(40,80)
\fmfcmd{style_def half_prop expr p =
    draw_plain p;
    shrink(.7);
        cfill (marrow (p, .5))
    endshrink;
	enddef;
}
\fmfstraight
\fmftopn{t}{3} \fmfbottom{b1}
\fmfv{d.shape=circle,d.filled=full,d.size=3thick}{i2}
\fmfv{d.shape=circle,d.filled=empty,d.size=3thick}{i1}
\fmf{half_prop}{b1,i1}
\fmf{phantom}{i2,t2}
\fmf{half_prop,right=0.5,tension=0.5}{i1,i2}
\fmf{half_prop,left=0.5,tension=0.5}{i1,i2}
\fmffreeze
\fmf{half_prop,right=0.3}{i1,t3}
\fmfv{label=\opC$^{11}(02,,31)$}{b1}
\end{fmfgraph*}
\end{fmffile}}
$\enspace + \enspace$
\parbox{40pt}{\begin{fmffile}{C11_5}
\begin{fmfgraph*}(40,80)
\fmfcmd{style_def half_prop expr p =
    draw_plain p;
    shrink(.7);
        cfill (marrow (p, .5))
    endshrink;
	enddef;
}
\fmfstraight
\fmftopn{t}{3} \fmfbottomn{b}{3}
\fmfv{d.shape=circle,d.filled=full,d.size=3thick}{i2}
\fmfv{d.shape=circle,d.filled=empty,d.size=3thick}{i1}
\fmf{phantom}{b2,i1}
\fmf{phantom}{i2,t2}
\fmf{half_prop}{i1,i2}
\fmffreeze
\fmf{half_prop,right=0.5,tension=0.5}{i1,i2}
\fmf{half_prop,left=0.5,tension=0.5}{i1,i2}
\fmf{half_prop,left=0.3}{b1,i2}
\fmf{half_prop,right=0.3}{i1,t3}
\fmfv{label=\opC$^{11}(04,,40)$}{b2}
\end{fmfgraph*}
\end{fmffile}}
$\enspace + \enspace$
\parbox{40pt}{\begin{fmffile}{C11_6}
\begin{fmfgraph*}(40,80)
\fmfcmd{style_def half_prop expr p =
    draw_plain p;
    shrink(.7);
        cfill (marrow (p, .5))
    endshrink;
	enddef;
}
\fmfstraight
\fmftop{t1} \fmfbottom{b1}
\fmfv{d.shape=circle,d.filled=full,d.size=3thick}{i2}
\fmfv{d.shape=circle,d.filled=empty,d.size=3thick}{i1}
\fmf{half_prop}{b1,i1}
\fmf{half_prop}{i2,t1}
\fmf{half_prop}{i1,i2}
\fmffreeze
\fmf{half_prop,right=0.5}{i1,i2}
\fmf{half_prop,left=0.5}{i1,i2}
\fmfv{label=\opC$^{11}(13,,31)$}{b1}
\end{fmfgraph*}
\end{fmffile}}
$- \enspace \Big[$
\parbox{10pt}{\fmfreuse{Bvertex}}
$\leftrightarrow$
\parbox{10pt}{\fmfreuse{Avertex}}
$\Big]$

\vspace{2.3\baselineskip}

\parbox{40pt}{\begin{fmffile}{C40}
\begin{fmfgraph*}(40,60)
\fmfcmd{style_def half_prop expr p =
    draw_plain p;
    shrink(.7);
        cfill (marrow (p, .5))
    endshrink;
	enddef;
}
\fmfcmd{
    	path quadrant, q[], otimes;
    	quadrant = (0, 0) -- (0.5, 0) & quartercircle & (0, 0.5) -- (0, 0);
    	for i=1 upto 4: q[i] = quadrant rotated (45 + 90*i); endfor
    	otimes = q[1] & q[2] & q[3] & q[4] -- cycle;
}
\fmfwizard
\fmfstraight
\fmftopn{t}{4} \fmfbottom{b1}
\fmfv{d.shape=otimes,d.filled=empty,d.size=3thick}{i1}
\fmf{phantom}{b1,i1}
\fmf{half_prop,tension=0.5}{i1,t1}
\fmf{half_prop,tension=0.5}{i1,t4}
\fmffreeze
\fmf{half_prop,tension=0.5}{i1,t2}
\fmf{half_prop,tension=0.5}{i1,t3}
\fmfv{label=\opC$^{40}$,label.d=0pt}{b1}
\end{fmfgraph*}
\end{fmffile}}
$ = \enspace$
\parbox{40pt}{\begin{fmffile}{C40_1}
\begin{fmfgraph*}(40,80)
\fmfcmd{style_def half_prop expr p =
    draw_plain p;
    shrink(.7);
        cfill (marrow (p, .5))
    endshrink;
	enddef;
}
\fmfstraight
\fmftopn{t}{7} \fmfbottom{b1}
\fmfv{d.shape=circle,d.filled=full,d.size=3thick}{i2}
\fmfv{d.shape=circle,d.filled=empty,d.size=3thick}{i1}
\fmf{phantom}{b1,i1}
\fmf{half_prop}{i2,t4}
\fmf{half_prop}{i1,i2}
\fmffreeze
\fmf{half_prop,left=0.3}{i1,t1}
\fmf{half_prop,left=0.3}{i1,t2}
\fmf{half_prop,left=0.3}{i1,t3}
\fmfv{label=\opC$^{40}(11,,40)$,label.d=5pt,l.angle=90}{b1}
\end{fmfgraph*}
\end{fmffile}}
$\enspace + \enspace$
\parbox{40pt}{\begin{fmffile}{C40_2}
\begin{fmfgraph*}(40,80)
\fmfcmd{style_def half_prop expr p =
    draw_plain p;
    shrink(.7);
        cfill (marrow (p, .5))
    endshrink;
	enddef;
}
\fmfstraight
\fmftopn{t}{5} \fmfbottom{b1}
\fmfv{d.shape=circle,d.filled=full,d.size=3thick}{i2}
\fmfv{d.shape=circle,d.filled=empty,d.size=3thick}{i1}
\fmf{phantom}{b1,i1}
\fmf{half_prop}{i2,t3}
\fmf{half_prop}{i1,i2}
\fmffreeze
\fmf{half_prop,left=0.3}{i1,t1}
\fmf{half_prop,left=0.3}{i2,t2}
\fmf{half_prop,right=0.3}{i2,t4}
\fmfv{label=\opC$^{40}(31,,20)$,label.d=5pt,l.angle=90}{b1}
\end{fmfgraph*}
\end{fmffile}}
$\enspace + \enspace$
\parbox{40pt}{\begin{fmffile}{C40_3}
\begin{fmfgraph*}(40,80)
\fmfcmd{style_def half_prop expr p =
    draw_plain p;
    shrink(.7);
        cfill (marrow (p, .5))
    endshrink;
	enddef;
}
\fmfstraight
\fmftopn{t}{6} \fmfbottom{b1}
\fmfv{d.shape=circle,d.filled=full,d.size=3thick}{i2}
\fmfv{d.shape=circle,d.filled=empty,d.size=3thick}{i1}
\fmf{phantom}{b1,i1}
\fmf{phantom}{i1,i2}
\fmf{half_prop,tension=0.5}{i2,t3}
\fmf{half_prop,tension=0.5}{i2,t4}
\fmffreeze
\fmf{half_prop,left=0.45}{i1,i2}
\fmf{half_prop,right=0.45}{i1,i2}
\fmf{half_prop,left=0.3}{i1,t1}
\fmf{half_prop,left=0.3}{i1,t2}
\fmfv{label=\opC$^{40}(22,,40)$,label.d=5pt,l.angle=90}{b1}
\end{fmfgraph*}
\end{fmffile}}
$- \enspace \Big[$
\parbox{10pt}{\fmfreuse{Bvertex}}
$\leftrightarrow$
\parbox{10pt}{\fmfreuse{Avertex}}
$\Big]$

\vspace{.8\baselineskip}

\parbox{40pt}{\begin{fmffile}{C31}
\begin{fmfgraph*}(40,60)
\fmfcmd{style_def half_prop expr p =
    draw_plain p;
    shrink(.7);
        cfill (marrow (p, .5))
    endshrink;
	enddef;
}
\fmfcmd{
    	path quadrant, q[], otimes;
    	quadrant = (0, 0) -- (0.5, 0) & quartercircle & (0, 0.5) -- (0, 0);
    	for i=1 upto 4: q[i] = quadrant rotated (45 + 90*i); endfor
    	otimes = q[1] & q[2] & q[3] & q[4] -- cycle;
}
\fmfwizard
\fmfstraight
\fmftopn{t}{3} \fmfbottom{b1}
\fmfv{d.shape=otimes,d.filled=empty,d.size=3thick}{i1}
\fmf{half_prop}{b1,i1}
\fmf{half_prop}{i1,t2}
\fmffreeze
\fmf{half_prop}{i1,t1}
\fmf{half_prop}{i1,t3}
\fmfv{label=\opC$^{31}$,l.dist=15pt}{b1}
\end{fmfgraph*}
\end{fmffile}}
$ = \enspace$
\parbox{40pt}{\begin{fmffile}{C31_1}
\begin{fmfgraph*}(40,80)
\fmfcmd{style_def half_prop expr p =
    draw_plain p;
    shrink(.7);
        cfill (marrow (p, .5))
    endshrink;
	enddef;
}
\fmfstraight
\fmftopn{t}{3} \fmfbottom{b1}
\fmfv{d.shape=circle,d.filled=full,d.size=3thick}{i2}
\fmfv{d.shape=circle,d.filled=empty,d.size=3thick}{i1}
\fmf{half_prop}{b1,i1}
\fmf{half_prop}{i2,t2}
\fmf{half_prop}{i1,i2}
\fmffreeze
\fmf{half_prop,left=0.3}{i2,t1}
\fmf{half_prop,right=0.3}{i2,t3}
\fmfv{label=\opC$^{31}(31,,11)$}{b1}
\end{fmfgraph*}
\end{fmffile}}
$\enspace + \enspace$
\parbox{40pt}{\begin{fmffile}{C31_2}
\begin{fmfgraph*}(40,80)
\fmfcmd{style_def half_prop expr p =
    draw_plain p;
    shrink(.7);
        cfill (marrow (p, .5))
    endshrink;
	enddef;
}
\fmfstraight
\fmftopn{t}{5} \fmfbottom{b1}
\fmfv{d.shape=circle,d.filled=full,d.size=3thick}{i2}
\fmfv{d.shape=circle,d.filled=empty,d.size=3thick}{i1}
\fmf{half_prop}{b1,i1}
\fmf{half_prop}{i2,t3}
\fmf{half_prop}{i1,i2}
\fmffreeze
\fmf{half_prop,left=0.3}{i1,t1}
\fmf{half_prop,left=0.3}{i1,t2}
\fmfv{label=\opC$^{31}(11,,31)$}{b1}
\end{fmfgraph*}
\end{fmffile}}
$\enspace + \enspace$
\parbox{40pt}{\begin{fmffile}{C31_3}
\begin{fmfgraph*}(40,80)
\fmfcmd{style_def half_prop expr p =
    draw_plain p;
    shrink(.7);
        cfill (marrow (p, .5))
    endshrink;
	enddef;
}
\fmfstraight
\fmftopn{t}{5} \fmfbottom{b1}
\fmfv{d.shape=circle,d.filled=full,d.size=3thick}{i2}
\fmfv{d.shape=circle,d.filled=empty,d.size=3thick}{i1}
\fmf{half_prop}{b1,i1}
\fmf{phantom}{i2,t3}
\fmf{half_prop,tension=0.5,left=0.5}{i1,i2}
\fmf{half_prop,tension=0.5,right=0.5}{i1,i2}
\fmffreeze
\fmf{half_prop}{i2,t2}
\fmf{half_prop}{i2,t4}
\fmf{half_prop,right=0.3}{i1,t5}
\fmfv{label=\opC$^{31}(22,,31)$}{b1}
\end{fmfgraph*}
\end{fmffile}}
$\enspace + \enspace$
\parbox{40pt}{\begin{fmffile}{C31_4}
\begin{fmfgraph*}(40,80)
\fmfcmd{style_def half_prop expr p =
    draw_plain p;
    shrink(.7);
        cfill (marrow (p, .5))
    endshrink;
	enddef;
}
\fmfstraight
\fmftopn{t}{5} \fmfbottomn{b}{3}
\fmfv{d.shape=circle,d.filled=full,d.size=3thick}{i2}
\fmfv{d.shape=circle,d.filled=empty,d.size=3thick}{i1}
\fmf{phantom}{b2,i1}
\fmf{half_prop}{i2,t3}
\fmf{half_prop,tension=0.5,left=0.5}{i1,i2}
\fmf{half_prop,tension=0.5,right=0.5}{i1,i2}
\fmffreeze
\fmf{half_prop,left=0.35}{i1,t1}
\fmf{half_prop,left=0.35}{i1,t2}
\fmf{half_prop,right=0.3}{b3,i2}
\fmfv{label=\opC$^{31}(13,,40)$}{b2}
\end{fmfgraph*}
\end{fmffile}}
$\enspace + \enspace$
\parbox{40pt}{\begin{fmffile}{C31_5}
\begin{fmfgraph*}(40,80)
\fmfcmd{style_def half_prop expr p =
    draw_plain p;
    shrink(.7);
        cfill (marrow (p, .5))
    endshrink;
	enddef;
}
\fmfstraight
\fmftopn{t}{5} \fmfbottomn{b}{3}
\fmfv{d.shape=circle,d.filled=full,d.size=3thick}{i2}
\fmfv{d.shape=circle,d.filled=empty,d.size=3thick}{i1}
\fmf{phantom}{b2,i1}
\fmf{phantom}{i2,t3}
\fmf{half_prop}{i1,i2}
\fmffreeze
\fmf{half_prop,left=0.3}{b1,i2}
\fmf{half_prop,right=0.3}{i1,t5}
\fmf{half_prop}{i2,t2}
\fmf{half_prop}{i2,t4}
\fmfv{label=\opC$^{31}(22,,20)$}{b2}
\end{fmfgraph*}
\end{fmffile}}
$\enspace + \enspace$
\parbox{40pt}{\begin{fmffile}{C31_6}
\begin{fmfgraph*}(40,80)
\fmfcmd{style_def half_prop expr p =
    draw_plain p;
    shrink(.7);
        cfill (marrow (p, .5))
    endshrink;
	enddef;
}
\fmfstraight
\fmftopn{t}{7} \fmfbottomn{b}{3}
\fmfv{d.shape=circle,d.filled=full,d.size=3thick}{i2}
\fmfv{d.shape=circle,d.filled=empty,d.size=3thick}{i1}
\fmf{phantom}{b2,i1}
\fmf{phantom}{i2,t4}
\fmf{half_prop}{i1,i2}
\fmffreeze
\fmf{half_prop,right=0.3}{b3,i2}
\fmf{half_prop,left=0.3}{i1,t1}
\fmf{half_prop,left=0.3}{i1,t2}
\fmf{half_prop,left=0.3}{i1,t3}
\fmfv{label=\opC$^{31}(02,,40)$}{b2}
\end{fmfgraph*}
\end{fmffile}}
$- \enspace \Big[$
\parbox{10pt}{\fmfreuse{Bvertex}}
$\leftrightarrow$
\parbox{10pt}{\fmfreuse{Avertex}}
$\Big]$

\vspace{2.3\baselineskip}

\parbox{40pt}{\begin{fmffile}{C22}
\begin{fmfgraph*}(40,60)
\fmfcmd{style_def half_prop expr p =
    draw_plain p;
    shrink(.7);
        cfill (marrow (p, .5))
    endshrink;
	enddef;
}
\fmfcmd{
    	path quadrant, q[], otimes;
    	quadrant = (0, 0) -- (0.5, 0) & quartercircle & (0, 0.5) -- (0, 0);
    	for i=1 upto 4: q[i] = quadrant rotated (45 + 90*i); endfor
    	otimes = q[1] & q[2] & q[3] & q[4] -- cycle;
}
\fmfwizard
\fmfstraight
\fmftopn{t}{2} \fmfbottomn{b}{3}
\fmfv{d.shape=otimes,d.filled=empty,d.size=3thick}{i1}
\fmf{half_prop}{b1,i1}
\fmf{half_prop}{i1,t2}
\fmf{half_prop}{i1,t1}
\fmf{half_prop}{b3,i1}
\fmfv{label=\opC$^{22}$,label.d=15pt}{b2}
\end{fmfgraph*}
\end{fmffile}}
$ = \enspace$
\parbox{40pt}{\begin{fmffile}{C22_1}
\begin{fmfgraph*}(40,80)
\fmfcmd{style_def half_prop expr p =
    draw_plain p;
    shrink(.7);
        cfill (marrow (p, .5))
    endshrink;
	enddef;
}
\fmfstraight
\fmftopn{t}{2} \fmfbottomn{b}{3}
\fmfv{d.shape=circle,d.filled=full,d.size=3thick}{i2}
\fmfv{d.shape=circle,d.filled=empty,d.size=3thick}{i1}
\fmf{half_prop}{b1,i1}
\fmf{half_prop}{b3,i1}
\fmf{half_prop,left=0.5}{i1,i2}
\fmf{half_prop,right=0.5}{i1,i2}
\fmf{half_prop}{i2,t2}
\fmf{half_prop}{i2,t1}
\fmfv{label=\opC$^{22}(22,,22)$}{b2}
\end{fmfgraph*}
\end{fmffile}}
$\enspace + \enspace$
\parbox{40pt}{\begin{fmffile}{C22_2}
\begin{fmfgraph*}(40,80)
\fmfcmd{style_def half_prop expr p =
    draw_plain p;
    shrink(.7);
        cfill (marrow (p, .5))
    endshrink;
	enddef;
}
\fmfstraight
\fmftopn{t}{4} \fmfbottomn{b}{5}
\fmfv{d.shape=circle,d.filled=full,d.size=3thick}{i2}
\fmfv{d.shape=circle,d.filled=empty,d.size=3thick}{i1}
\fmf{phantom,tension=2}{b3,i1}
\fmf{half_prop,left=0.4}{i1,i2}
\fmf{half_prop,right=0.4}{i1,i2}
\fmf{phantom}{i2,t2}
\fmf{phantom}{i2,t3}
\fmffreeze
\fmf{half_prop,right=0.4}{b5,i2}
\fmf{half_prop,right=0.4}{b4,i2}
\fmf{half_prop,left=0.4}{i1,t1}
\fmf{half_prop,left=0.4}{i1,t2}
\fmfv{label=\opC$^{22}(04,,40)$}{b3}
\end{fmfgraph*}
\end{fmffile}}
$\enspace + \enspace$
\parbox{40pt}{\begin{fmffile}{C22_3}
\begin{fmfgraph*}(40,80)
\fmfcmd{style_def half_prop expr p =
    draw_plain p;
    shrink(.7);
        cfill (marrow (p, .5))
    endshrink;
	enddef;
}
\fmfstraight
\fmftopn{t}{3} \fmfbottomn{b}{3}
\fmfv{d.shape=circle,d.filled=full,d.size=3thick}{i2}
\fmfv{d.shape=circle,d.filled=empty,d.size=3thick}{i1}
\fmf{half_prop}{b2,i1}
\fmf{half_prop,tension=0.5,left=0.5}{i1,i2}
\fmf{half_prop,tension=0.5,right=0.5}{i1,i2}
\fmf{half_prop}{i2,t2}
\fmffreeze
\fmf{half_prop,left=0.3}{b1,i2}
\fmf{half_prop,right=0.3}{i1,t3}
\fmfv{label=\opC$^{22}(13,,31)$}{b2}
\end{fmfgraph*}
\end{fmffile}}
$\enspace + \enspace$
\parbox{40pt}{\begin{fmffile}{C22_4}
\begin{fmfgraph*}(40,80)
\fmfcmd{style_def half_prop expr p =
    draw_plain p;
    shrink(.7);
        cfill (marrow (p, .5))
    endshrink;
	enddef;
}
\fmfstraight
\fmftopn{t}{3} \fmfbottomn{b}{3}
\fmfv{d.shape=circle,d.filled=full,d.size=3thick}{i2}
\fmfv{d.shape=circle,d.filled=empty,d.size=3thick}{i1}
\fmf{half_prop,tension=0.5}{b1,i1}
\fmf{half_prop,tension=0.5}{b3,i1}
\fmf{half_prop}{i1,i2}
\fmf{half_prop}{i2,t2}
\fmffreeze
\fmf{half_prop,right=0.3}{i1,t3}
\fmfv{label=\opC$^{22}(11,,22)$}{b2}
\end{fmfgraph*}
\end{fmffile}}
$\enspace + \enspace$
\parbox{40pt}{\begin{fmffile}{C22_5}
\begin{fmfgraph*}(40,80)
\fmfcmd{style_def half_prop expr p =
    draw_plain p;
    shrink(.7);
        cfill (marrow (p, .5))
    endshrink;
	enddef;
}
\fmfstraight
\fmftopn{t}{2} \fmfbottomn{b}{3}
\fmfv{d.shape=circle,d.filled=full,d.size=3thick}{i2}
\fmfv{d.shape=circle,d.filled=empty,d.size=3thick}{i1}
\fmf{half_prop}{b2,i1}
\fmf{half_prop}{i1,i2}
\fmf{half_prop,tension=0.5}{i2,t1}
\fmf{half_prop,tension=0.5}{i2,t2}
\fmffreeze
\fmf{half_prop,right=0.3}{b3,i2}
\fmfv{label=\opC$^{22}(22,,11)$}{b2}
\end{fmfgraph*}
\end{fmffile}}
$\enspace + \enspace$
\parbox{40pt}{\begin{fmffile}{C22_6}
\begin{fmfgraph*}(40,80)
\fmfcmd{style_def half_prop expr p =
    draw_plain p;
    shrink(.7);
        cfill (marrow (p, .5))
    endshrink;
	enddef;
}
\fmfstraight
\fmftopn{t}{5} \fmfbottomn{b}{3}
\fmfv{d.shape=circle,d.filled=full,d.size=3thick}{i2}
\fmfv{d.shape=circle,d.filled=empty,d.size=3thick}{i1}
\fmf{half_prop}{b2,i1}
\fmf{half_prop}{i1,i2}
\fmf{phantom}{i2,t3}
\fmffreeze
\fmf{half_prop,left=0.3}{i1,t1}
\fmf{half_prop,left=0.3}{i1,t2}
\fmf{half_prop,right=0.3}{b3,i2}
\fmfv{label=\opC$^{22}(02,,31)$}{b2}
\end{fmfgraph*}
\end{fmffile}}
$\enspace + \enspace$
\parbox{40pt}{\begin{fmffile}{C22_7}
\begin{fmfgraph*}(40,80)
\fmfcmd{style_def half_prop expr p =
    draw_plain p;
    shrink(.7);
        cfill (marrow (p, .5))
    endshrink;
	enddef;
}
\fmfstraight
\fmftopn{t}{3} \fmfbottomn{b}{5}
\fmfv{d.shape=circle,d.filled=full,d.size=3thick}{i2}
\fmfv{d.shape=circle,d.filled=empty,d.size=3thick}{i1}
\fmf{phantom}{b3,i1}
\fmf{half_prop}{i1,i2}
\fmf{half_prop}{i2,t2}
\fmffreeze
\fmf{half_prop,left=0.3}{b1,i2}
\fmf{half_prop,left=0.3}{b2,i2}
\fmf{half_prop,right=0.3}{i1,t3}
\fmfv{label=\opC$^{22}(13,,20)$}{b3}
\end{fmfgraph*}
\end{fmffile}}
$- \enspace \Big[$
\parbox{10pt}{\fmfreuse{Bvertex}}
$\leftrightarrow$
\parbox{10pt}{\fmfreuse{Avertex}}
$\Big]$

\vspace{2\baselineskip}

\caption{\label{f:BIMSRG2flow} Diagrams contributing to the $(2,2;2)$ commutator under the hypothesis that $\opC$ is Hermitian or anti-Hermitian, i.e., only the subset $\{\opC^{ij}, i\geq j\}$ is explicitly computed.}
\end{figure*}

\subsection{BIMSRG(2)}
\label{BIMSRG2diag}

The working-horse of standard IMSRG calculations is the IMSRG(2) approximation. Let us now formulate the corresponding approximation in BIMSRG. In either the direct integration of the flow equation for the grand potential or the Magnus expansion, the BIMSRG(2) truncation scheme is defined by the (repeated) use of the $(N_{\opA}, N_{\opB}; N_{\opC})=(2,2;2)$ approximation to the elementary commutator under the hypothesis that $\opC$ is Hermitian/anti-Hermitian. 

The diagrams contributing to $\{\opC^{ij}, i\geq j\}$ in this approximation are displayed in Fig.~\ref{f:BIMSRG2flow}. They correspond to all diagrams with topologies characterized by $d_\mathrm{max} \leq 2$. Allowing the exchange $[\opA \leftrightarrow \opB]$ and exploiting the (anti-)Hermiticity of $\opC$, these 28 diagrams actually account for a total of 82 diagrams. 

The algebraic expressions of the BIMSRG(2) flow equations are obtained by applying the algebraic rules to the diagrams displayed in Fig.~\ref{f:BIMSRG2flow} and by identifying 
\begin{align}
\opA &\rightarrow \eta(s) \, ,\notag \\
\opB &\rightarrow \Omega(s)  \,, \\
\opC &\rightarrow \dds{\opno{\Omega}{}{}}(s) \, , \notag
\end{align}
for the case of direct integration or
\begin{align}
\opA &\rightarrow \magnus(s) \, ,\notag \\
\opB &\rightarrow \text{ad}_\magnus^{(l-1)}(\eta) \,, \\
\opC &\rightarrow \text{ad}_\magnus^{(l)}(\eta) \, , \notag
\end{align}
in the case of the Magnus expansion.
The evaluated expressions are collected in~\ref{BIMSRG2approx}.

\subsection{BIMSRG(n)}

While the BIMSRG(2) truncation is an efficient workhorse, one eventually wishes to push to BIMSRG(3) in order to obtain high-accuracy results. The numerical code described below allows one to obtain the corresponding equations in a matter of seconds. The additional 82 diagrams\footnote{Performing the exchange $[\opA \leftrightarrow \opB]$ and exploiting the Hermitian/anti-Hermitian character of $\opC$, these extra 82 diagrams actually account for a total of 264 diagrams.} arising in the $(N_{\opA}, N_{\opB}; N_{\opC})=(3,3;3)$ approximation are too numerous and the equations too lengthy to be displayed in this article; the interested reader is invited to run the \textbf{\texttt{ADG}} code accordingly. 

For orientation, Tab.~\ref{t:bimsrg_n} lists the number of diagrams arising from the commutator in the $(N_{\opA}, N_{\opB}; N_{\opC})=(n,n;n)$ approximation up to $n=9$, i.e.~all diagrams with topologies characterized by $d_\mathrm{max} = n$ up to $n=9$ (BIMSRG(9)). Of course, the number of diagrams increases rapidly. The restriction obtained by virtue of the (anti-)Hermiticity of $\opC$ can go up to a factor of 2 in the ideal case where the number of $\opC^{kk}$ terms is very small with respect to other diagrams. Together with using the symmetry in the truncation, this yields an ideal reduction by a factor of 4 in the number of distinct diagrams that have to be considered explicitly.

\begin{table*}
\begin{center}
\begin{tabular}{|c|c|c|c|c|c|c|c|c|c|}
\hline
BIMSRG(n) & 1 & 2 & 3 & 4 & 5 & 6 & 7 & 8 & 9 \\ \hline \hline
Naive counting & 10 & 72 & 264 & 700 & 1550 & 2930 & 5152 & 8424 & 13046 \\ \hline
Using $N_\opA = N_\opB$ & 5 & 36 & 132 & 350 & 775 & 1465 & 2576 & 4212 & 6523 \\ \hline
Using $N_\opA = N_\opB$ and Hermiticity & 4 & 24 & 82 & 208 & 452 & 830 & 1436 & 2320 & 3558 \\ \hline
\end{tabular}
\end{center}
\caption{
Number of new diagrams appearing at each order of the BIMSRG(n) truncation. The naive counting refers to computing explicitly all diagrams contributing to both $+\opA\opB$ and  $-\opB\opA$. The growing number of diagrams can be reduced by using the symmetry in the truncation and the Hermiticity of \opC.}
\label{t:bimsrg_n}
\end{table*}

\section{\textbf{\texttt{ADG}} code}
\label{ADGcode}

Given the set of diagrammatic rules, all diagrams contributing to  $\opC=[\opA,\opB]$ in the $(N_\opA,N_\opB;N_\opC)$ scheme are automatically produced and evaluated via the \textbf{\texttt{ADG}} code. The values of $N_\opA$ and $N_\opB$ do not have to be the same whereas $N_\opC$ can be chosen to take any value between 0 and $N_\opA+N_\opB-1$; i.e. the code operates is not limited to the specific constraints of the BIMSRG(n) truncation scheme.

The code generates expressions according to the following algorithms.

\subsection{Diagram generation}

\begin{enumerate}
    \item Select either $\opA$ or $\opB$ to be the top vertex and associate $\opB$ (resp.~$\opA$) with the bottom vertex.
    \begin{enumerate}
        \item Select a valid vertex degree $d_\opC$ in $\{0,2,\dots,2N_\opC\}$ stipulating the total number of external lines of the output operator.
        \begin{enumerate}
            \item Partition $d_\opC$ between incoming and outgoing external lines, i.e.~pick integers $i$ and $j$ to select $\opC^{ij}$ such that $d_{\opC^{ij}} \equiv i+j = d_\opC$.
            \begin{enumerate}
                \item Connect the external legs to $\opA$ and $\opB$ vertices in all possible ways. This must be done while keeping the number of these connections strictly smaller than $2N_\opA$ and $2N_\opB$, respectively, in order to allow for at least one contraction between $\opA$ and $\opB$. Additionally, the numbers of external legs connected to vertex $\opA$ and to vertex $\opB$ have to be either both even or both odd, restricting the number of configurations.
                \item Add $n_l$ internal lines connecting vertex $\opA$ to vertex $\opB$ in all possible ways. The possible values of $n_l$ must be such that Eq.~\eqref{identity2} is fulfilled with the degrees $d_{\opA^{kl}}$ and $d_{\opB^{mn}}$ of the involved vertices $\opA^{kl}$ and $\opB^{mn}$ being in $\{2,4,\dots,2N_\opA\}$ and $\{2,4,\dots,2N_\opB\}$, respectively.
            \end{enumerate}
            \item Go back to i.~and exhaust all valid partitions.
        \end{enumerate}
        \item Go back to (a) and exhaust all valid vertex degrees $d_\opC$.
    \end{enumerate}
    \item Go back to 1.~and repeat the operation with permuted vertices.
\end{enumerate}

\subsection{Diagram evaluation}

\begin{enumerate}
    \item If the top vertex corresponds to the operator $\opB$, insert a minus sign as the diagram is associated to the $-\opB\opA$ term of the commutator.
    \item If incoming (resp.~outgoing) external legs are tied to both $\opA$ and $\opB$ vertices, write a permutation operator $P$ associated to both groups of lines. Permutations are to be taken between the group of lines entering (resp.~leaving) vertex $\opA$ and the group entering (resp.~leaving) vertex $\opB$.
    \item Write a symmetry factor $1/(n_l!)$ where $n_l$ is the number of internal lines connecting the two vertices.
    \item For each vertex, write the matrix element of the corresponding operator with the quasi-particle labels of the attached out-going and incoming lines in canonical order.
    \item Add the factor $(-1)^{n_c}$ where $n_c$ denotes the number of lines crossing.
    \item Write a summation over the quasi-particle labels associated to the internal lines.
\end{enumerate}

\subsection{Output of the program}

The typical output associated to a BIMSRG diagram in the code is\footnote{All BIMSRG diagrams generated by \textbf{\texttt{ADG}} are made in such a way that both their graphical representation and algebraic expression are in canonical form. Additionally, there is a direct correspondence between a diagram expression and its drawing as obtained by the code.}

\paragraph{Diagram 56 ($-BA$):}
\begin{equation*}
C^{22}(40,04) = - \frac{1}{2}\sum_{p_{1}p_{2}} B^{04}_{k_{3}k_{4} p_{1}p_{2}} A^{40}_{p_{1}p_{2} k_{1}k_{2}}
\end{equation*}
\begin{center}
\parbox{40pt}
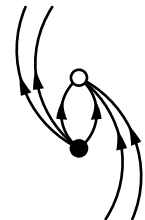
{\begin{fmffile}{diag_55}
\begin{fmfgraph*}(40,80)
\fmfcmd{style_def half_prop expr p =
draw_plain p;
shrink(.7);
	cfill (marrow (p, .5))
endshrink;
enddef;}
\fmfstraight
\fmftopn{t}{5}\fmfbottomn{b}{5}
\fmf{phantom}{b3,v1}
\fmf{phantom}{v1,v2}
\fmf{phantom}{v2,t3}
\fmfv{d.shape=circle,d.filled=full,d.size=3thick}{v1}
\fmfv{d.shape=circle,d.filled=empty,d.size=3thick}{v2}
\fmffreeze
\fmf{half_prop,right=0.5}{v1,v2}
\fmf{half_prop,left=0.5}{v1,v2}
\fmf{half_prop,right=0.4}{b5,v2}
\fmf{half_prop,right=0.4}{b4,v2}
\fmf{half_prop,left=0.4}{v1,t1}
\fmf{half_prop,left=0.4}{v1,t2}
\end{fmfgraph*}
\end{fmffile}}
\end{center}

\section{Use of the \texttt{ADG} program}
\label{programuse}

\texttt{ADG} has been designed to work on any computer with a Python3 distribution, and successfully tested on 
common Linux
distributions as well as MacOS. In addition to the Python base, the \emph{setuptools} and \emph{distutils} packages must already be installed, which is the case in most current distributions. Having \emph{pip} installed eases the process but is not technically required. The \emph{NumPy}, \emph{NetworkX} and \emph{SciPy} libraries are automatically downloaded during the installation process. Additionally, one needs a \LaTeX\ distribution installed with the PDF\LaTeX\ compiler for \texttt{ADG} to produce a pdf file from the output, if so desired.

\subsection{Installation}

\subsubsection{From the Python Package Index}

The easiest way to install \texttt{ADG} is to obtain it from the Python Package Index\footnote{\url{https://pypi.org/project/adg/}} by entering the following command
\begin{verbatim}
pip3 install adg
\end{verbatim}
Provided \emph{setuptools} is already installed, \emph{pip} takes care of downloading and installing \texttt{ADG} as well as \emph{NumPy}, \emph{SciPy} and \emph{NetworkX}. Once a new version of \texttt{ADG} is released, one can install it by entering the command
\begin{verbatim}
pip3 install --upgrade adg
\end{verbatim}

\subsubsection{From the source files}

Once the \texttt{ADG} source files are downloaded from the GitHub repository\footnote{\url{https://github.com/adgproject/adg}}, one must enter the project folder and either run
\begin{verbatim}
pip3 install .
\end{verbatim}
or
\begin{verbatim}
python3 setup.py install
\end{verbatim}
With this method, \emph{pip}\footnote{Depending on the system, it might be necessary either to use the "--user" flag to install it only for a specific user or to run the previous command with "sudo -H" to install it system-wide.} also takes care of downloading and installing \emph{NumPy}, \emph{NetworkX} and \emph{SciPy}.

\subsection{Run the program}

\subsubsection{Batch mode}

The most convenient way to use \texttt{ADG} is to run it in batch mode with the appropriate flags. For example, to run the program and generate BIMSRG diagrams at order 3, i.e.~using $N_A = N_B = N_C = 3$, one can use
\begin{verbatim}
adg -o 3 -t BIMSRG -d -c
\end{verbatim}
where the \texttt{-o} flag is for the order, \texttt{-t} for the type of theory, \texttt{-d} indicates that the diagrams must be drawn and \texttt{-c} that \texttt{ADG} must compile the \LaTeX\ output. The program supports less conventional truncation schemes as well. Two integers after the \texttt{-o} flag will be interpreted as settings for $N_\opA = N_\opB$ and $N_\opC$ respectively, and three integers as $N_\opA$, $N_\opB$ and $N_\opC$ respectively.
A complete list of the command-line options can be found in the program's documentation (see Sec.~\ref{documentation}) or by typing
\begin{verbatim}
adg -h
\end{verbatim}

Currently, \texttt{ADG} supports HF-MBPT by using \texttt{-t MBPT}, straight BMBPT by using \texttt{-t BMBPT}, off-diagonal BMBPT by using \texttt{-t PBMBPT} and BIMSRG by using \texttt{-t BIMSRG}. Though the algorithms described in the previous sections can be used regardless of the truncation order, \texttt{ADG} has been arbitrarily restricted to order 10 or less to avoid major overloads of the system due to rapidly growing number of expressions or diagrams. Users are advised to first launch calculations at low orders (2, 3 or 4 typically) to develop an intuition for the time and memory needs for specific applications.

\subsubsection{Interactive mode}

As an alternative to the batch mode, \texttt{ADG} can be run on a terminal by entering the command
\begin{verbatim}
adg -i
\end{verbatim}
A set of questions must be answered using the keyboard to configure and launch the calculation. The interactive mode then proceeds identically to the batch mode.

\subsection{Steps of a program run}

Here, we briefly describe the main steps of a typical \texttt{ADG} run:
\begin{itemize}
\item Select options by using the command-line flags or keyboard input by the user in an interactive session.
\item \texttt{ADG} creates a list of adjacency matrices for the appropriate theory and order using \emph{NumPy}, and feeds them to \emph{NetworkX}, which creates \emph{MultiDiGraph} objects.
\item If needed, checks are performed on the list of graphs to remove topologically equivalent or ill-defined graphs.
\item The list of topologically unique graphs is used to produce \emph{Diagram} objects that store the graph as well as some of its associated properties depending on the theory (HF status, excitation level, etc.). The expressions associated to the graphs are eventually extracted.
\item The program prints on the terminal the number of diagrams per category and writes the \LaTeX\ output file, the details of which depend on the options selected by the user, as well as a list of adjacency matrices associated to the diagrams. Other output files may be produced, depending on the theory and the user's input.
\item If asked by the user, the program performs the PDF\LaTeX\ compilation.
\item Unnecessary temporary files are removed and the programs exits.
\end{itemize}

\subsection{Documentation}
\label{documentation}

\subsubsection{Local documentation}

Once the source files have been downloaded, a quick start guide is available in the \texttt{README.md} file. Once \texttt{ADG} is installed, it is possible to read its manpages through
\begin{verbatim}
man adg
\end{verbatim}
or a brief description of the program and its options through
\begin{verbatim}
adg -h
\end{verbatim}
A more detailed HTML documentation can be generated directly from the source files by going into the \texttt{docs} directory and running
\begin{verbatim}
make html
\end{verbatim}
The documentation is then stored in \texttt{docs/build/html}, with the main file being \texttt{index.html}. A list of other supported documentation formats is available by running
\begin{verbatim}
make help
\end{verbatim}

\subsubsection{Online documentation}

The full HTML documentation is available online under \url{https://github.com/adgproject/adg}  and support for the program can be obtained by reporting bugs or other issues on the GitHub repository's tracker at \url{https://github.com/adgproject/adg}.

\section{Conclusion and Outlook}
\label{sec:outlook}

A novel single-reference expansion many-body method appropriate to the \textit{ab initio} description of superfluid nuclei is formulated in terms of a particle-number-breaking Bogoliubov vacuum: The Bogoliubov in-medium similarity renormalization group (BIMSRG) formalism. Such an extension of the standard, i.e., symmetry-conserving, single-reference IMSRG approach parallels those developed recently within the framework of coupled cluster, self-consistent Green's function and many-body perturbation theories. Furthermore, BIMSRG complements the already existing multi-reference in-medium similarity renormalization group (MR-IMSRG) formalism that is also applicable to open-shell systems.

The derivation of BIMSRG equations relies on the computation of the commutator between two operators of potentially high ranks. To perform such a computation efficiently, a diagrammatic method is designed on the basis of a normal-ordered form of the operators with respect to the Bogoliubov vacuum. While low-rank truncations of the commutator can easily be worked out manually on the basis of the diagrammatic rules, it  becomes impractical and error-prone to work out higher orders, e.g. BIMSRG(3) equations, in this way. To overcome this limitation, we introduce the third version (v3.0.0) of the \textbf{\texttt{ADG}} code, which can  automatically (1) generate all valid BIMSRG(n) diagrams and (2) evaluate their algebraic expressions in a matter of seconds. This is achieved in such a way that equations can easily be retrieved for both the flow equation and the Magnus expansion variants of BIMSRG.

Based on this work, the first objective is to perform full-fledged \textit{ab initio} calculations of singly open-shell nuclei at the BIMSRG(2) level by employing a spherical implementation that also makes use of angular-momentum coupling tools~\cite{Tichai2020amc}. Later on, an extension to BIMSRG(3) can be envisioned. Another future direction is to extend the \textbf{\texttt{ADG}} code~\cite{Arthuis2018adg1,Arthuis2020adg2} to automatically derive equations at play within MR-IMSRG, in particular to systematically and safely access those defining the MR-IMSRG(3) approximation.

\section*{Acknowledgements}
The authors thank S.~Bogner and A.~Schwenk for useful discussions.
This publication is based on work supported in part by the Deutsche  Forschungsgemeinschaft  (DFG,  German Research Foundation) – Projektnummer 279384907 – SFB 1245, the Max Planck Society, by the BMBF Contract No.~05P18RDFN1, and the National Science Foundation under Grants No. PHY-1614130.

\section*{Data Availability Statement}
The datasets generated during and/or analysed during the current study are available
in the GitHub repository, 
\href{https://github.com/adgproject/adg}{\nolinkurl{https://github.com/adgproject/adg}}.

\begin{appendix}
\setcounter{equation}{0}
\renewcommand\theequation{A.\arabic{equation}}
\allowdisplaybreaks

\section{BIMSRG(2)}
\label{BIMSRG2approx}

\subsection{BIMSRG(2) flow equation}

Following the strategy laid out in Sec.~\ref{BIMSRG2diag}, the BIMSRG(2) flow equations for the normal-ordered pieces of the grand potential read
\begin{subequations}
\label{eq:omegaflow}
\begin{align}
\dds{\opno{\Omega}{00}{}}=& 
\frac{1}{2} \sum_{pq} \opno{\eta}{02}{pq} \opno{\Omega}{20}{pq} \nonumber \\
&+ \frac{1}{4!}\sum_{pqrs} \opno{\eta}{04}{pqrs} \opno{\Omega}{40}{pqrs}  \nonumber \\
& - \exchange \, , \label{eq:omega00flow}
\\
\dds{\opno{\Omega}{20}{k_1 k_2}}=& 
\permtwo{k_1}{k_2} \sum_p 
\opno{\eta}{11}{k_2 p} \opno{\Omega}{20}{k_1 p} \nonumber \\
&
+ \frac{1}{2}\sum_{pq}  \opno{\eta}{22}{k_1 k_2 pq} \opno{\Omega}{20}{pq}  \nonumber \\
&
+ \frac{1}{2} \sum_{pq}  \opno{\eta}{02}{pq}  \opno{\Omega}{40}{k_1 k_2 pq} \nonumber \\
&+ \permtwo{k_1}{k_2} \frac{1}{3!}\sum_{pqr}   \opno{\eta}{13}{k_2 pqr} \opno{\Omega}{40}{k_1 pqr}  \nonumber \\
&- \exchange\, , \label{eq:omega20flow} \\
\dds{\opno{\Omega}{11}{k_1 k_2} } =& 
\sum_p \opno{\eta}{11}{k_1 p} \opno{\Omega}{11}{p k_2}  \nonumber \\
&+ \sum_p  \opno{\eta}{02}{k_2 p} \opno{\Omega}{20}{p k_1}  \nonumber \\
&+\frac{1}{2} \sum_{pq}   \opno{\eta}{13}{k_1 k_2 pq} \opno{\Omega}{20}{pq}  \nonumber \\
&+\frac{1}{2} \sum_{pq}   \opno{\eta}{02}{pq} \opno{\Omega}{31}{pq k_1 k_2}  \nonumber \\
&+ \frac{1}{3!} \sum_{pqr} \opno{\eta}{04}{k_2 pqr} \opno{\Omega}{40}{pqr k_1} \nonumber \\
&+ \frac{1}{3!} \sum_{pqr} \opno{\eta}{13}{k_1 pqr} \opno{\Omega}{31}{pqr k_2} \nonumber \\
& - \exchange\, , \label{eq:omega11flow} \\
\dds{\opno{\Omega}{40}{k_1k_2k_3 k_4} } =& 
\permtwo{k_1k_2k_3}{k_4}  \sum_{p}  \opno{\eta}{11}{k_4 p} \opno{\Omega}{40}{k_1k_2k_3 p}    \nonumber \\
&+ \permtwo{k_1}{k_2k_3k_4} \sum_p  \opno{\eta}{31}{k_2 k_3 k_4 p} \opno{\Omega}{20}{k_1 p}  \nonumber \\
&+  \permtwo{k_1 k_2}{k_3 k_4} \frac{1}{2} \sum_{pq}  \opno{\eta}{22}{k_3 k_4 pq}  \opno{\Omega}{40}{k_1 k_2 pq}  \nonumber \\
& - \exchange  \, , \label{eq:omega40flow} \\
\dds{\opno{\Omega}{31}{k_1k_2k_3 k_4}} =& 
\permtwo{k_1 k_2}{k_3} \sum_{p}  \opno{\eta}{11}{k_3 p}  \opno{\Omega}{31}{k_1 k_2 p k_4 } \nonumber \\
&+\permtwo{k_1k_2}{k_3} \sum_{p}  \opno{\eta}{22}{k_1 k_2 k_4 p} \opno{\Omega}{20}{p k_4}  \nonumber \\
&+ \sum_p   \opno{\eta}{02}{pk_4} \opno{\Omega}{40}{k_1 k_2 k_3 p}  \nonumber \\
&+ \sum_p  \opno{\eta}{31}{k_1 k_2 k_3 p} \opno{\Omega}{11}{pk_4}  \nonumber \\
&+ \permtwo{k_1 k_2}{k_3} \frac{1}{2} \sum_{pq}  \opno{\eta}{22}{k_1 k_2 pq}  \opno{\Omega}{31}{pq k_3 k_4}  \nonumber \\
&+ \permtwo{k_1 k_2}{k_3} \frac{1}{2} \sum_{pq}  \opno{\eta}{13}{k_3 pq k_4}  \opno{\Omega}{40}{k_1 k_2 pq }  \nonumber \\
&- \exchange \, ,  \label{eq:omega31flow} \\
\dds{\opno{\Omega}{22}{k_1k_2k_3 k_4}} =& 
\frac{1}{2} \sum_{pq} \opno{\eta}{04}{pq k_3 k_4} \opno{\Omega}{40}{k_1 k_2 p q} \nonumber \\
& +\permtwo{k_1}{k_2} \permtwo{k_3}{k_4} \frac{1}{2} \sum_{pq}  \opno{\eta}{13}{k_1 k_3 p q}  \opno{\Omega}{31}{pq k_2 k_4} \nonumber \\
&+ \frac{1}{2} \sum_{pq} \opno{\eta}{22}{k_1 k_2 p q}  \opno{\Omega}{22}{pq k_3 k_4}  \nonumber \\
&+ \permtwo{k_1}{k_2}  \sum_p  \opno{\eta}{11}{k_1 p} \opno{\Omega}{22}{p k_2 k_3 k_4 }  \nonumber \\
&+ \permtwo{k_3}{k_4} \sum_p  \opno{\eta}{22}{k_1 k_2 p k_4} \opno{\Omega}{11}{pk_3} \nonumber \\
&+ \permtwo{k_1}{k_2}   \sum_p  \opno{\eta}{13}{k_1 k_3 k_4 p} \opno{\Omega}{20}{pk_2}  \nonumber \\
&+ \permtwo{k_3}{k_4} \sum_p  \opno{\eta}{02}{p k_4}  \opno{\Omega}{31}{k_1 k_2 p k_3} \nonumber \\
&- \exchange \, , \label{eq:omega22flow}
\end{align}%
\end{subequations}
where the required permutation operators are defined through
\begin{subequations}
\label{permutoperators}
\begin{align}
    \permtwo{k_1}{k_2} &\equiv 1 - P_{k_1k_2}\, , \\
    \permtwo{k_1}{k_2k_3} &\equiv 1 - P_{k_1k_2} - P_{k_1 k_3}\, , \\
    \permtwo{k_1}{k_2k_3k_4} &\equiv 1 - P_{k_1k_2} - P_{k_1 k_3} - P_{k_1 k_4}\, , \\
    \permtwo{k_1k_2}{k_3k_4} &\equiv 1 - P_{k_1k_3} - P_{k_1 k_4} - P_{k_2k_3} - P_{k_2 k_4} + P_{k_1k_3} P_{k_2 k_4}\, . 
\end{align}
\end{subequations}
The shorthand-notation $\exchange$ indicates that additional contributions are to be obtained by exchanging $\eta$ and $\Omega$ in all previous terms. The specific (anti-)symmetry properties of the matrix elements of $(\text{d}\Omega^{ij}/\text{d}s)^{ij}$ can be used to only compute them for a subset of quasi-particle indices.

Note that even when a canonical HFB vacuum is employed at the beginning of the flow, i.e., $\opno{\Omega}{20}{k_1k_2}(0) = \opno{\Omega}{02}{k_1k_2}(0) =0$, the corresponding evolved matrix elements are non-zero due to RG-corrections from contracted normal-ordered two-body components.

\subsection{BIMSRG(2) Magnus expansion}

The Magnus expansion relies on the computation of the $l$-fold nested commutator  $\text{ad}_\magnus^{(l)}(\eta)$ defined recursively in Eq.~\eqref{adjoint}. At each step of the nesting, the BIMSRG(2) corresponds to setting
\begin{equation}
\opA \rightarrow M(s) \, / \,  \opB \rightarrow \text{ad}_\magnus^{(l-1)}(\eta)  \, / \,  \opC \rightarrow \text{ad}_\magnus^{(l)}(\eta) \, .
\end{equation}
in the $(N_{\opA}, N_{\opB}; N_{\opC})=(2,2;2)$ approximation to the elementary commutator. The corresponding equations could be easily obtained by adapting Eq.~\eqref{eq:omegaflow}. This procedure is to be repeated iteratively until the maximum value of $l$ is reached.

\subsection{Computational scaling}

The computational scaling of the BIMSRG tensor network is driven by the commutator involving the highest particle rank, e.g., in the case of BIMSRG(2)
\begin{align}
    [ \opno{\eta}{[4]}{} , \opno{\Omega}{[4]}{}]^{[4]} \, .
    \label{eq:commscal}
\end{align}
Equation~\eqref{eq:commscal} requires $N^6$ evaluations, where $N$ characterizes the size of the quasi-particle basis or, equivalently, the dimension of the one-body Hilbert space.
In standard IMSRG(2), the scaling is $N^6$ as well, but certain types of generators allow further reductions. For the frequently employed White and imaginary-time generators, the computational effort is driven by the evaluation of particle-particle ladder diagrams~\cite{Herg16PR}. It scales as $n_p^4 n_h^2$ where $n_p$ and $n_h$ denote the number of particle and hole states, respectively, similar to Coupled Cluster with singles and doubles (CCSD). Since $n_p \gg n_h$, the quasi-particle formulation involves a significantly higher cost even though the scaling is bound to be polynomial in $N$ at any finite truncation order. As such, the evaluation of BIMSRG(2) is comparable to the solution of the BCC equations at the singles and doubles level (BCCSD) once appropriate intermediates are defined~\cite{Sign14BogCC}.

\subsection{Perturbation theory diagnostics}

The decoupling of the Bogoliubov reference state from its  quasi-particle excitations may be quantified during the flow in terms of various tensor norms of operator matrix elements. The repeated evaluation of low-order MBPT energy corrections was also successfully used as a diagnostic in previous IMSRG studies. Since BIMSRG(2) resums all diagrams up to third order, second- and third-order grand-potential corrections must eventually vanish at convergence. This property can thus be used as a diagnostic tool. 

Limiting the flowing grand potential to $\Omega^{[4]}(s)$ terms, the Rayleigh-Schrodinger partition is introduced as
\begin{align}
    \Omega(s) = \Omega_0(s) + \Omega_1(s) \, ,
\end{align}
with
\begin{subequations}
\begin{align}
    \Omega_0(s) &\equiv \opno{\Omega}{00}{}(s) + \opno{\bar \Omega}{11}{}(s) \, , \\
    \Omega_1(s) &\equiv \opno{\Omega}{20}{}(s) + \opno{\Omega}{02}{}(s) + \opno{\breve \Omega}{11}{}(s) + \opno{\Omega}{40}{}(s) + \opno{\Omega}{31}{}(s) \notag \\ &\phantom{=}
    + \opno{\Omega}{22}{}(s) + \opno{\Omega}{13}{}(s) + \opno{\Omega}{04}{}(s)\, , 
\end{align}
\end{subequations}
and where the splitting
\begin{align}
    \opno{\Omega}{11}{}(s) \equiv \opno{\bar \Omega}{11}{}(s) + \opno{\breve \Omega}{11}{}(s)
\end{align}
is introduced such that the unperturbed part is defined as a diagonal one-body operator.
Low-order corrections to the ground-state grand potential are evaluated via BMBPT~\cite{Tichai18BMBPT,Tichai2020review}. For example, the second-order\footnote{The numbering of the orders used here corresponds to the usual convention among MBPT practitioners, at variance with what is used in the \texttt{ADG} code~\cite{Arthuis2018adg1} mentioned below.} reads as
\begin{align}
\text{E}^{(2)}_0(s) -\lambda \text{A}^{(2)}_0(s) &\equiv \frac{1}{4} \sum_{pq} \frac{ \opno{\Omega}{20}{pq}(s) \opno{\Omega}{02}{pq}(s) }{ \opno{\bar \Omega}{11}{pp}(s) + \opno{\bar \Omega}{11}{qq}(s) }  \\ 
& -\frac{1}{24} \sum_{pqrs} \frac{ \opno{\Omega}{40}{pqrs}(s) \opno{\Omega}{04}{pqrs}(s) }{ \opno{\bar \Omega}{11}{pp}(s) + \opno{\bar \Omega}{11}{qq}(s) + \opno{\bar \Omega}{11}{rr}(s) + \opno{\bar \Omega}{11}{ss}(s)} \, . \notag
\end{align}
Even starting from a canonical HFB vacuum, non-canonical diagrams need to be evaluated consistently during the flow since $\opno{\Omega}{20,02}{pq}(s) \neq 0$ for $s>0$. Canonical and non-canonical third-order diagrams are additionally needed and are conveniently obtained using \texttt{ADG}~\cite{Arthuis2018adg1}.

\subsection{Generators}
\label{sec:generators}

So far, no specific form of the generator has been assumed. In the following, generators appropriate to BIMSRG(2) calculations are derived. Given that the flowing grand potential is limited to $\Omega^{[4]}(s)$ terms, it is sufficient to drive the couplings to two and four quasi-particle excitations
\begin{subequations}
\label{decouplingBIMSRG(2)}
\begin{align}
\la \Phi | \Omega(s) | \Phi^{pq} \ra &= \opno{\Omega}{20}{pq}(s) \, \\
\la \Phi | \Omega(s) | \Phi^{pqrs} \ra &= \opno{\Omega}{40}{pqrs}(s) \, ,
\end{align}
\end{subequations}
to zero as $s \rightarrow \infty$. To do so, one must employ generators of the form
\begin{align}
    \eta(s) \equiv \opno{\eta}{20}{}(s) + \opno{\eta}{02}{}(s) + 
    \opno{\eta}{40}{}(s) + \opno{\eta}{04}{}(s) \, ,
\end{align}
with matrix elements defined by
\begin{subequations}
\begin{align}
    \opno{\eta}{20}{pq}(s) &\equiv \opno{G}{}{pq} \opno{\Omega}{20}{pq}(s) \, , \\
    \opno{\eta}{02}{pq}(s) &\equiv \opno{G}{}{pq} \opno{\Omega}{02}{pq}(s) \, , \\
    \opno{\eta}{40}{pqrs}(s) &\equiv \opno{G}{}{pqrs} \opno{\Omega}{40}{pqrs}(s) \, , \\
    \opno{\eta}{04}{pqrs}(s) &\equiv \opno{G}{}{pqrs} \opno{\Omega}{04}{pqrs}(s) \, .
\end{align}
\label{eq:genansatz}
\end{subequations}
In Eq.~\eqref{eq:genansatz}, the tensor $G$ has been included to enforce anti-Hermiticity of $\eta$, i.e.,
\begin{subequations}
\begin{align}
    \opno{G}{}{pq} &=-\opno{G}{}{qp} \, , \\
    \opno{G}{}{pqrs} &=- \opno{G}{}{rspq} \, .
\end{align}
\end{subequations}
Specific choices for $\eta$ relate to fixing $G$. 

In order to analyse the performance of the above ansatz for the generator, a perturbative analysis~\cite{Herg16PR} of the flow equation can be performed on the basis of the so-called Epstein-Nesbet (EN) partitioning of the grand potential
\begin{align}
    \opno{\Omega}{}{}(s) = \opno{\tilde{\Omega}}{}{0}(s) + \opno{\tilde{\Omega}}{}{1}(s)
\end{align}
where
\begin{subequations}
\begin{align}
    \opno{\tilde{\Omega}}{}{0}(s) \equiv& \opno{\Omega}{00}{}(s) + \sum_{p} \opno{\Omega}{11}{pp}(s) \, \beta^\dagger_{p} \beta_{p} \nonumber \\
    & + \frac{1}{4} \sum_{pq} \opno{\Omega}{22}{pqpq}(s) \, \beta^\dagger_{p} \beta^\dagger_{q} \beta_{q}\beta_{p}\, , \\
    \opno{\tilde{\Omega}}{}{1}(s) \equiv& \sum_{p\neq q} \opno{\Omega}{11}{pq}(s) \, \beta^\dagger_{p} \beta_{q} +\frac{1}{4} \sum_{\substack{pqrs \\ (pq) \neq (rs)}} \opno{\Omega}{22}{pqrs}(s) \,
    \beta^\dagger_{p} \beta^\dagger_{q} \beta_{s}\beta_{r}  \nonumber \\
    &+ \opno{\Omega}{40}{}(s) + \opno{\Omega}{31}{}(s) + \opno{\Omega}{13}{}(s) + \opno{\Omega}{04}{}(s)
    \, , 
\end{align}
\end{subequations}
such that diagonal matrix elements of $\opno{\Omega}{22}{}(s)$ have been included into the unperturbed grand potential $\ptorder{0}$. Counting insertions of $\opno{\tilde{\Omega}}{}{1}(s)$ ($\opno{\tilde{\Omega}}{}{0}(s)$) as order $\ptorder{1}$ ($\ptorder{0}$), the perturbative expansion of Eqs.~\eqref{eq:omega20flow} and \eqref{eq:omega40flow} yields
\begin{subequations}
\label{perturbativeODEs}
\begin{align}
\dds{} \opno{\Omega}{20}{pq}
&= - (\opno{\Omega}{11}{pp} + \opno{\Omega}{11}{qq})
\opno{\eta}{20}{pq} -
\opno{\Omega}{22}{pqpq} \opno{\eta}{20}{pq}  + \ptorder{2} \notag \\
& \equiv -\opno{\Delta}{}{pq}\opno{\eta}{20}{pq}  + \ptorder{2} \, , \\
\dds{} \opno{\Omega}{40}{pqrs}
    &= - (\opno{\Omega}{11}{pp} + \opno{\Omega}{11}{qq} + 
    \opno{\Omega}{11}{rr} + \opno{\Omega}{11}{ss}) \opno{\eta}{40}{pqrs} \notag \\
    &\phantom{=}-(\opno{\Omega}{22}{pqpq} +
    \opno{\Omega}{22}{prpr} +
    \opno{\Omega}{22}{psps} \notag \\ &\phantom{=}  +
    \opno{\Omega}{22}{qrqr} +
    \opno{\Omega}{22}{qsqs} +
    \opno{\Omega}{22}{rsrs}) \opno{\eta}{40}{pqrs} +  \ptorder{2}  \notag \\
    &\equiv - \opno{\Delta}{}{pqrs} \opno{\eta}{40}{pqrs} + \ptorder{2} \, ,
\end{align}
\end{subequations}
where
\begin{subequations}
\begin{align}
    \opno{\Delta}{}{pq}(s) &\equiv
    \la \Phi^{pq} | \Omega(s) | \Phi^{pq} \ra - \la \Phi | \Omega(s) | \Phi\ra \nonumber \\
    &= \opno{\Omega}{11}{pp}(s) + \opno{\Omega}{11}{qq}(s) + \opno{\Omega}{22}{pqpq}(s) \, , \\
    \opno{\Delta}{}{pqrs}(s) &\equiv
    \la \Phi^{pqrs} | \Omega(s) | \Phi^{pqrs} \ra - \la \Phi | \Omega(s) | \Phi\ra  \nonumber \\
    &= \opno{\Omega}{11}{pp}(s) + \opno{\Omega}{11}{qq}(s) + 
    \opno{\Omega}{11}{rr}(s) + \opno{\Omega}{11}{ss}(s) \nonumber \\
    &\phantom{=}+ \opno{\Omega}{22}{pqpq}(s) + \opno{\Omega}{22}{prpr}(s)+ \opno{\Omega}{22}{psps}(s) \nonumber \\
    &\phantom{=} + \opno{\Omega}{22}{qrqr}(s) + \opno{\Omega}{22}{qsqs}(s) + \opno{\Omega}{22}{rsrs}(s) \, ,
\end{align}
\end{subequations}
define EN energy denominators that are fully symmetric under the permutation of any pair of quasi-particle indices. Inserting Eq.~\eqref{eq:genansatz} into Eq.~\eqref{perturbativeODEs}, one obtains 
\begin{subequations}
\begin{align}
    \dds{} \opno{\Omega}{20}{pq} &= 
    - \opno{\Delta}{}{pq} \opno{G}{}{pq} \opno{\Omega}{20}{pq}
    + \ptorder{2} \, , \\
    \dds{} \opno{\Omega}{40}{pqrs} &= 
    - \opno{\Delta}{}{pqrs} \opno{G}{}{pqrs} \opno{\Omega}{40}{pqrs}
    + \ptorder{2} \, ,
\end{align}
\end{subequations}
which can be formally integrated while ignoring the $s$ dependence of the EN energy denominators as
\begin{subequations}
\label{expsuppression}
\begin{align}
    \opno{\Omega}{20}{pq}(s) &= 
    \opno{\Omega}{20}{pq}(0) e^{-s\opno{\Delta}{}{pq} \opno{G}{}{pq}} \, , \\
    \opno{\Omega}{40}{pqrs}(s) &= 
    \opno{\Omega}{40}{pqrs}(0) e^{-s\opno{\Delta}{}{pqrs} \opno{G}{}{pqrs}} \, .
\end{align}
\end{subequations}
Under the assumption that $\opno{\Delta}{}{pq}$ and $\opno{\Delta}{}{pqrs}$ are positive\footnote{To ensure the suppression in Eq.~\eqref{expsuppression}, one can slightly modify Eq.~\eqref{eq:genansatz} according to
\begin{subequations}
\begin{align}
    \opno{\eta}{20}{pq} &\equiv \text{sgn}( \opno{\Delta}{}{pq}) \opno{G}{}{pq} \opno{\Omega}{20}{pq} \, , \\
    \opno{\eta}{02}{pq} &\equiv \text{sgn}( \opno{\Delta}{}{pq}) \opno{G}{}{pq} \opno{\Omega}{02}{pq} \, , \\
    \opno{\eta}{40}{pqrs} &\equiv \text{sgn}( \opno{\Delta}{}{pqrs}) \opno{G}{}{pqrs} \opno{\Omega}{40}{pqrs} \, , \\
    \opno{\eta}{04}{pqrs} &\equiv \text{sgn}( \opno{\Delta}{}{pqrs}) \opno{G}{}{pqrs} \opno{\Omega}{04}{pqrs} \, .
\end{align}
\end{subequations}}, $\opno{\Omega}{20}{pq}(s)$ and $\opno{\Omega}{40}{pqrs}(s)$ are exponentially suppressed such that the decoupling invoked in Eq.~\eqref{eq:decoupling} is indeed realized for the BIMSRG(2) approximation when $s \rightarrow \infty$. It is to be noted that the rate at which the matrix elements  $\opno{\Omega}{20}{pq}(s)$ and $\opno{\Omega}{40}{pqrs}(s)$ are suppressed increases with the size of the associated EN denominator $|\opno{\Delta}{}{pq}|$ and $|\opno{\Delta}{}{pqrs}|$, as is customary for an RG evolution.

\end{appendix}

\bibliographystyle{apsrev4-2}

\bibliography{strongint}

\end{document}